\documentclass[10pt]{article}
\usepackage{hyperref}
\usepackage{float}
\usepackage{amssymb,amsmath,graphicx,latexsym,cite}

\begin{document}

\title{A Cosmological Holographic Reconstruction of $f(Q)$ Theory}

\author{Pameli Saha \thanks{pameli.saha15@gmail.com}~~ Prabir Rudra \thanks{prudra.math@gmail.com (Corresponding author)}\\
\emph{\small{$^*$ Department of Mathematics, Amity University, Maharashtra-410206, India.}}\\
\emph{\small{$^\dagger$ Department of Mathematics, Asutosh
College, Kolkata-700 026, India.}}}

%\author{Prabir Rudra \thanks{prudra.math@gmail.com, rudra@associates.iucaa.in}\\\emph{\small{Department of Mathematics, Asutosh College,
%Kolkata-700 026, India.}}}

\maketitle

\begin{abstract}
This paper explores a cosmological reconstruction scheme in the
background of $f(Q)$ gravity theory from a Holographic
perspective. The basic motivation for this work is that the
reconstruction is performed from a holographic origin, which has
its roots in the black hole thermodynamics and quantum gravity.
Dark energy models inspired by holographic prescription are used
to reconstruct the $f(Q)$ gravity models. Two such models, namely
the Granda-Oliveros holographic dark energy model and its
generalization, the Chen-Jing model are considered for the study.
Different scale factors are used and a thorough reconstruction
scheme is set up using the dark energy models. The observationally
constrained values of the free model parameters have been used to
form the reconstructed models. Finally, a thorough investigation
of the energy conditions has been performed to check the
cosmological viability of the reconstructed $f(Q)$ models. As an
outcome, we get some very promising and cosmologically viable
$f(Q)$ models that present some interesting properties and demand
further investigation. Finally a method is discussed how the
constructed $f(Q)$ models can be reconciled with a generalized
holographic dark energy.
\end{abstract}



%%%%%%%%%%%%%%%%%%%%%%%%
\section{Introduction}
%%%%%%%%%%%%%%%%%%%%%%%%
The late-time acceleration of the universe has been a hotly
debated issue for the past two decades. This feature of the cosmos
is confirmed by recent developments in observational cosmology,
such as the type Ia supernovae \cite{sn1, sn2, sn3}, Cosmic
Microwave Background Radiation \cite{cmb1}, and large-scale
structure observations \cite{cmb2, cmb3, cmb4}. However, the true
problem is in the theoretical framework being developed to explain
this universe-processing mechanism. It is commonly recognized that
general relativity (GR) remains the most successful fundamental
gravitational theory to describe the large-scale structure of the
cosmos for over a century. In the context of cosmology, the FLRW
spacetime and the matter source provide precise answers for the
scale factor $a(t)$, which aids in our understanding of the
universe's expansion. There are two types of accelerated expansion
that the cosmos must go through to regain FLRW spacetime. One of
them is the dynamics of early time inflation, which may be
examined by adding a scalar field to the Einstein-Hilbert action.
Additionally, the Einstein field equation can be made to account
for the current accelerating expansion by using a cosmological
constant. This relatively simple model, known as the $\Lambda$CDM
model, fits all the available observations that were previously
presented and explains the universe's late-time cosmic
acceleration. However, issues with vacuum energy scale \cite{vc1,
vc2} fine-tuning negatively impact this lovely model.
Consequently, it is critical to look at alternate strategies or
generalizations of basic theories of gravity.

The so-called $f(R)$ gravity \cite{fr1, fr2}, which extends the
Hilbert-Einstein action to a general function of the Ricci scalar
$R$, is the most basic generalization of general relativity. This
modified theory of gravity is well recognized for its ability to
replicate the entire cosmological history and the behavior of the
cosmological constant $\Lambda$ \cite{lamb1, lamb2, lamb3}, as
well as for its successful description of cosmic acceleration.
Several approaches have been used in the literature to recover the
properties of $\Lambda$CDM using modified theories of gravity.
Further modifications to $f(R)$ theories led to the development of
$f(R,T)$ \cite{frt1} and $f(R,T^2)$ \cite{frt2, frt3} theories
where $T$ is the trace of the energy-momentum tensor $T_{\mu\nu}$
and $T^2=T_{\mu\nu}T^{\mu\nu}$. In these theories, a non-zero
coupling between the matter and the geometry sectors is considered
which gives rise to interesting phenomenologies. The underlying
Riemannian geometry \cite{riemann1} (formulated in the Riemann
metrical space) at the core of all these classical theories,
including GR, is a common property shared by all the
aforementioned extensions of GR. Given the incompatibility of
these theories at certain scales, it makes sense to consider
whether some of the contradictions that have dogged these
classical theories over time might be eliminated if the underlying
geometry could be substituted with a far more universal geometric
framework. Weyl \cite{weyl1} made such a bold attempt, with the
primary goal being the geometrical unification of electromagnetic
and gravity. We are aware that the Levi-Civita connection is the
fundamental tool for comparing vector lengths in Riemannian
geometry and that it is compatible with the metric. Weyl's theory
employs a completely new method that uses two connections, one of
which is in charge of a vector's direction during parallel transit
and the other of which carries its length information. The
electromagnetic potential is the physical counterpart of the
length link. The non-zero covariant divergence of the metric
tensor, which gives rise to a new geometrical quantity known as
non-metricity $Q$, is the theory's most remarkable characteristic.
Weyl's geometry is a masterwork of mathematics with an equally
intricate physical structure.

In literature, we see that there are two basic formulations of
gravity, namely the curvature ($R\neq 0$, $\tau=0$) and the
teleparallel ($R=0$, $\tau\neq 0$) formulation, where $\tau$ is
the scalar torsion. However, the non-metricity $Q$ disappears in
these two formulations. In parallel transport, the variation in a
vector's length is represented geometrically by $Q$. Currently, a
non-vanishing non-metricity $Q$ was thought to be the fundamental
geometrical variable in charge of all gravitational interaction
types in a third equivalent formalism of general relativity.
Symmetric teleparallel gravity (STG) is the name given to this
theory \cite{q1}. In this instance, the energy-momentum density is
represented by the Einstein pseudotensor, which ultimately
transforms into a genuine tensor in the geometric representation.
Subsequent investigations led to the extension of the STG to
$f(Q)$ gravity \cite{q2}, which is sometimes referred to as
non-metric gravity and coincident general relativity. In \cite{q3,
q4}, the cosmology of $f(Q)$ gravity and its empirical limitations
were examined. The STG framework has seen several improvements
during the last few decades \cite{q5, q6, q7, q8, q9}. The authors
of ref.\cite{q9} have suggested expanding the $f(Q)$ gravity by
taking into account the non-minimal coupling that exists between
the matter Lagrangian $L_m$ and the non-metricity $Q$. The
non-conservation of the energy-momentum tensor and the
introduction of an additional force in the geodesic equation of
motion are quite predictable outcomes of the non-minimal coupling
between the geometry and matter sectors. Another generalization of
the theory, the $f(Q,T)$ gravity, was proposed by Xu et al. in
\cite{q10}. In this case, the gravity Lagrangian is essentially an
arbitrary function of $Q$ and the trace of the energy-momentum
tensor $T$. The model's cosmic evolution was examined and the
field equations were determined.

The entropy of a system is determined by its area rather than its
volume, according to the holographic principle, which has its
roots in black hole thermodynamics \cite{14, 15}. This holographic
concept, which also has linkages to string theory \cite{14, 19},
has led to the development of holographic dark energy (HDE)
\cite{16, 17, 18}. An ultraviolet cutoff, or the furthest distance
allowed by the framework, is known to be associated with a quantum
field theory \cite{20}. The vacuum energy, a type of dark energy
with a holographic origin, is directly connected to this
ultraviolet cutoff. The reader might consult \cite{21} for a
comprehensive overview of HDE. HDE has been the subject of much
research, both in its basic and expanded forms, and the model has
proven to be quite effective throughout time \cite{22, 23, 24, 25,
26, 27, 28}. The compatibility of HDE models with observational
data has been one of their main success factors \cite{29, 30, 31}.

The goal of the cosmological reconstruction approach is to use
updated gravity theories to precisely recover the $\Lambda$CDM
features and determine the universe's expansion history. The
intricacy of field equations, which makes it challenging to get an
exact and numerical solution that can be compared with
observations, complicates studies of the physics of such theories.
Using the reconstruction technique, which is predicated on the
accurate understanding of the universe's expansion history, one
inverts the field equations to ascertain which class of modified
theory gives rise to a given flat FRW model. A lot of research has
been performed in this direction of late. For example, Nojiri et
al. \cite{recon1} developed an intriguing plan for the
cosmological reconstruction of $f(R)$ gravity in terms of
e-folding. In $f(R)$ gravity, Dunsby et al. \cite{recon2} found
that an accurate reconstruction of $\Lambda$CDM development
requires an additional degree of freedom for the matter
components. Since many popular $f(G)$ models \cite{fg1} do not
allow for any accurate power-law solutions, Goheer et al.
\cite{fg2} showed that exact power-law solutions in $f(G)$ gravity
exist only for a certain class of models. In \cite{tuhina1} the
authors have performed a reconstruction scheme for the $f(\tau,
T)$ Lagrangian for various cosmological scenarios.

In this work, we are interested in studying a cosmological
reconstruction scheme in the background of $f(Q)$ gravity using a
direct correspondence from the holographic dark energy. Since
holographic dark energy has its origin in black hole
thermodynamics, any reconstruction scheme using this form of dark
energy is expected to produce new and interesting results.
Moreover, we have seen that reconstruction schemes using other
forms of dark energy models have been quite successful and is
widely available in the literature. Moreover, any modified gravity
model reconstructed from HDE is expected to give viable results
both in the early and late universe (which is a salient feature of
this reconstruction study). The connection of the holographic dark
energy with quantum gravity adds to the interest. Herein lies our
basic motivation to undertake this study. The paper is organized
as below: In section II we revisit the basic field equations of
$f(Q)$ gravity. In section III, we discuss the holographic dark
energy models to be used in this study. In section IV we perform
the reconstruction analysis. Then in section V, we check the
energy conditions for the reconstructed models. In section VI we
study a reconstruction scheme using the generalized HDE. Finally,
the paper ends with some discussion and concluding remarks in
section VII.

\section{$f(Q)$ Gravity}
In this section we will discuss the basic concepts and the related
equations of $f(Q)$ gravity. We consider the following action for
$f(Q)$ gravity,
\begin{equation}\label{action}
S=\int
\left[-\frac{1}{2\kappa^{2}}f(Q)+\mathcal{L}_{m}\right]\sqrt{-g}d^{4}x
\end{equation}
where $\mathcal{L}_{m}$ is the matter Lagrangian, $f(Q)$ is an
arbitrary function of the non-metricity scalar $Q$ and $g$ is the
determinant of the metric tensor $g_{\mu\nu}$. We have the
non-metricity scalar as
\begin{equation}\label{nm1}
Q=-\frac{1}{4}Q_{\alpha\beta\gamma}Q^{\alpha\beta\gamma}+\frac{1}{2}Q_{\alpha\beta\gamma}Q^{\gamma\beta\alpha}+\frac{1}{4}Q_{\alpha}Q^{\alpha}-\frac{1}{2}Q_{\alpha}\Tilde{Q}^{\alpha},
\end{equation}
where
\begin{equation}\label{nm2}
Q_{\alpha}\equiv Q_{\alpha\mu}^{\mu},
\end{equation}
\begin{equation}\label{nm3}
\Tilde{Q}^{\alpha}\equiv Q_{\mu}^{\mu\alpha}
\end{equation}
and the non-metricity tensor as,
\begin{equation}\label{nm4}
Q_{\alpha\mu\nu}\equiv\nabla_{\alpha}g_{\mu\nu}.
\end{equation}
If we take $f(Q)=Q$ then we get the Symmetric Teleparallel
Equivalent of General Relativity. Now from the equations
(\ref{action}),(\ref{nm1}), (\ref{nm2}), (\ref{nm3}), (\ref{nm4}),
the field equations are generated as
\begin{eqnarray*}\label{field1}
\frac{2}{\sqrt{-g}}\nabla_{\alpha}\Big\{\sqrt{-g}g_{\beta\nu}f_{Q}\Big[-\frac{1}{2}L^{\alpha\mu\beta}+\frac{1}{4}g^{\mu\beta}(Q^{\alpha}-\Tilde{Q}^{\alpha})-\frac{1}{8}(g^{\alpha\mu}Q^{\beta}+g^{\alpha\beta}Q^{\mu})\Big]\Big\}
\end{eqnarray*}
\begin{equation}
+f_{Q}\Big[-\frac{1}{2}L^{\mu\alpha\beta}-\frac{1}{8}(g^{\mu\alpha}Q^{\beta}+g^{\mu\beta}Q^{\alpha})+\frac{1}{4}g^{\alpha\beta}(Q^{\mu}-\Tilde{Q}^{\mu})\Big]Q_{\nu\alpha\beta}+\frac{1}{2}\delta^{\mu}_{\nu}f=\kappa^{2}T^{\mu}_{\nu},
\end{equation}
where $f_{Q}\equiv\frac{\partial f}{\partial Q}$. Here the
deformation tensor is given by
\begin{equation}\label{deform1}
L^{\alpha}_{\mu\nu}=\frac{1}{2}Q^{\alpha}_{\mu\nu}-Q_{\mu
\nu}^{\alpha}
\end{equation}
and the matter energy-momentum tensor is
\begin{equation}\label{emt1}
T_{\mu\nu}=-\frac{2}{\sqrt{-g}}\frac{\delta(\sqrt{-g}\mathcal{L}_{m}}{\delta
g^{\mu\nu}}
\end{equation}

Next, we consider the homogeneous and isotropic
Friedmann-Lemaitre-Robertson-Walker (FLRW) metric,
\begin{equation}
ds^{2}=-dt^{2}+a^{2}(t)(dx^{2}+dy^{2}+dz^{2}),
\end{equation}
where $a(t)$ is the scale factor with the cosmic time $t$. Now,
the field equations (\ref{field1}) of $f(Q)$ gravity (where
$Q=6H^{2}$) in the framework of FLRW spacetime reduce to
\begin{equation}\label{frw1}
3H^{2}=\kappa^{2}(\rho_{m}+\rho_{de}),
\end{equation}
\begin{equation}\label{frw2}
3H^{2}+\dot{H}=-\kappa^{2}(p_{m}+p_{de})
\end{equation}
where $H$ is the Hubble parameter given by $H=\frac{\dot{a}}{a}$.
The dark energy density and pressure in the above FLRW equations
are given by
\begin{equation}\label{rhode1}
\rho_{de}=\frac{3}{\kappa^{2}}\Big[H^{2}(1-2f_{Q})+\frac{f}{6}\Big],
\end{equation}
\begin{equation}\label{pde1}
p_{de}=-\frac{1}{\kappa^{2}}\Big[2\dot{H}(1-f_{Q})+\frac{f}{2}+3H^{2}(1-8f_{QQ}\dot{H}-2f_{Q}\Big].
\end{equation}
Here $\rho_{m}$ is the energy density of matter and $p_{m}$ is the
pressure of the matter fluid, and a dot represents a derivative
with respect to the cosmic time $t$. The conservation equation of
matter component is given as
\begin{equation}
\dot{\rho}_{m}+3H(\rho_{m}+p_{m})=0.
\end{equation}

\section{Dark Energy Models}
The observational dataset according to (i) Hubble Telescope, (ii)
Large Scale Structure (LSS), (iii) Supernovae Ia, Weak Lensing,
(iv) Baryon Acoustic Oscillations (BAO) and (v) Cosmic Microwave
Background (CMB) radiation stipulate the evidence of 70\% of the
entire universe by Dark Energy and the rest part dominated by Dark
matter and Ordinary matter. The $f(Q)$ gravity is a modified form
of gravity that can be considered equivalent to a dark energy
model if the strong energy condition is violated i.e.,
$\rho_{de}+3p_{de}<0$ implying
\begin{equation}
24f_{QQ}\dot{H}H^{2}-f<6\dot{H}(1-f_{Q})
\end{equation}
This relation helps us to investigate the capacity of $f(Q)$
models to act as dark energy. The modified Friedmann equation is
defined as
\begin{equation}
8\pi G\rho=\Sigma_{\Sigma n_{i}}A(X, Y, ...)\frac{\partial^{n}f(X,
Y, ...)}{\partial X^{n_{1}}\partial Y^{n_{2}} ...}.
\end{equation}
where the action $Q$ plays as the matter part for $f(X, Y, ...)$. For example, the form $\rho=\rho(H, \dot{H}, ...)$ of Holographic dark energy can be written as $\rho=\rho(X, Y, ...)$ to solve some partial differential equations of $f(X, Y, ...)$ to procure the physical nature of dark energy for some cosmogenic model.\\

One of the most perplexing problems in contemporary cosmology is
dark energy. Specifically, dark energy might even be a quantum
gravity problem. In this sense, the dark energy of our universe is
largely described by the holographic principle, which is one of
the tenets of quantum gravity. The covariant generalized version
of the Holographic dark energy has been proposed to correspond
with the modified gravity to explore the dark energy scenario in
the inflationary phase in ref. \cite{Nojiri}. In
ref.\cite{Nojiri1} also the authors have described the generalized
holographic dark energy to express the inflationary expansion of
the early universe. The Granda and Oliveros (GO) model has been
described to prospect the evolution of dark energy density and the
deceleration parameter in connection with a holographic origin
\cite{Karami}. Granda-Oliveros cut-off is written as
\cite{Koussour}
\begin{equation}
L_{GO}=(\alpha H^{2}+\beta\dot{H})^{-\frac{1}{2}}
\end{equation}
where the Hubble parameter $H$ is $\frac{\dot{a}}{a}$ and 'dot'
denotes the derivative with respect to the cosmic time $t$. This
model involves two parameters $\alpha$ and $\beta$. Since the
expansion of the universe is attributed to the presence of dark
energy, it is logical to consider its density dependence on the
rate of change of the Hubble parameter ($H$). So we can take the
dark energy density as a function of $H$ and $\dot{H}$. Such a
dependence is controlled by the model parameters $\alpha$ and
$\beta$. If the IR cut-off of holographic dark energy is taken as
particle horizon then it can not explain the accelerated expansion
of our present universe. However, if we take the IR cut-off as the
future event horizon, then another problem arises in the
description of our universe. In this work, we will work with the
Granda-Oliveros model \cite{Nojiri1, Koussour} as a model for dark
energy. We have $L_{GO}\propto R$ as the limiting case $\{\alpha,
\beta\}=\{2, 1\}$. The best fit observational data
\cite{Malekjani} for a non-flat universe is,
\begin{equation}
\alpha=0.8824^{+0.2180}_{-0.1163}(1\sigma)^{+0.2213}_{-0.1378}(2\sigma),~~~~~~~~~\beta=0.5016^{+0.0973}_{-0.0871}(1\sigma)^{+0.1247}_{-0.1102}(2\sigma),
\end{equation}
and for a flat universe is \cite{Malekjani}
\begin{equation}
\alpha=0.8502^{+0.0984}_{-0.0875}(1\sigma)^{+0.1299}_{-0.1064}(2\sigma),~~~~~~~~~~\beta=0.4817^{+0.0842}_{-0.0773}(1\sigma)^{+0.1176}_{-0.0955}(2\sigma).
\end{equation}
The energy densities for the Granda-Oliveros model can be written
as \cite{Malekjani}
\begin{equation}\label{go2}
\rho_{D}=3c^{2}(\alpha H^{2}+\beta\dot{H}),
\end{equation}
where $\alpha$ and $\beta$ are the dimensionless parameters. A
generalized version of the Granda-Oliveros model, dubbed as the
Chen-Jing model is available in the literature. It's energy
density is defined as a function of $H$, $\dot{H}$ and $\ddot{H}$
\cite{Chen, Ray}
\begin{equation}\label{cg1}
\rho_{D}=3c^{2}\Big[\alpha\Big(\frac{\ddot{H}}{H}\Big)+\beta\dot{H}+\gamma
H^{2}\Big]
\end{equation}
where $\alpha$, $\beta$ and $\gamma$ are the dimensionless
parameters. If $\alpha=0$ then Chen-Jing model (\ref{cg1}) reduces
to Granda-Oliveros model (\ref{go2}) \cite{Granda, Sheykhi}.
Again, if $\alpha=0$, $\beta=1$ and $\gamma=2$ then the energy
density with IR cut-off will be proportional to the reciprocal of
the square root of scalar curvature i.e., $L\propto
\frac{1}{\sqrt{R}}$. For the two models (\ref{go2}) and
(\ref{cg1}), $c$ is a dimensionless parameter of unit order.

In this paper, we will explore the reconstruction schemes of
$f(Q)$ gravity using these two Holographic dark energy models
given in eqns. (\ref{go2}) and (\ref{cg1}) under the different
types of scale factor evolution.

%%%%%%%%%%%%%%%%%%%%%%%%%%%%%%%%%%
\section{The Reconstruction scheme}
%%%%%%%%%%%%%%%%%%%%%%%%%%%%%%%%%%
In this section, drawing motivation from \cite{recon2, fg1, fg2,
tuhina1, revv1, revv2} we study the reconstruction scheme of
$f(Q)$ gravity model with the different types of scale factors.
Here, we consider two dark energy models (i) Granda-Oliveros and
(ii) Chen-Jing to correspond with $f(Q)$ gravity. Observationally
constrained values of dimensionless parameters $\alpha$, $\beta$,
and $\gamma$ will be considered to study the physical nature of
our proposed model. Below, we consider five different scale
factors for our reconstruction scheme. It should be noted that the
basic motivation for this work is that the reconstruction is
performed from a holographic origin, which has its roots in the
black hole thermodynamics and quantum gravity.

\subsection{Case I} Here we consider the power law form of scale factor \cite{Anagnostopoulos} for the evolution of the universe given by
\begin{equation}\label{scale1}
a(t)=a_{0}t^{m},
\end{equation}
where $m>0$ is a constant and $a_{0}$ denotes the present value of
$a(t)$. This form of scale factor gives solutions for the horizon
problem, the flatness problem for the late universe, and the
problem coming from the early universe \cite{Abreu, Karananas}.
This form of scale factor has been taken in \cite{Amani, Odintsov,
Shamir, Pasqua1} to analyze different types of modified gravity
theories to describe the late universe. These types of research
motivate us to investigate our proposed model under this scale
factor. Using equations (\ref{rhode1}), (\ref{go2}), and
(\ref{scale1}) we get a differential equation of $f(Q)$ as
\begin{equation}\label{difeq1}
2Qf'(Q)-f(Q)=\{1-\frac{c^{2}\kappa^{2}}{m}(\alpha m-\beta)\}Q
\end{equation}
The general solution of equation (\ref{difeq1}) for a non-flat
universe is given by,
\begin{equation}\label{Sol1}
f(Q)=c_{1}\sqrt{Q}+\{1-\frac{c^{2}\kappa^{2}}{m}(0.8824m-0.5016)\}Q
\end{equation}
and for a flat universe, we have,
\begin{equation}\label{Sol2}
f(Q)=c_{1}\sqrt{Q}+\{1-\frac{c^{2}\kappa^{2}}{m}(0.8502m-0.4817)\}Q
\end{equation}
$c_{1}$ is an arbitrary constant in the above solutions. Solving
equations (\ref{rhode1}), (\ref{cg1}) and (\ref{scale1}) we get a
differential equation of $f(Q)$ for the Chen-Jing model as
\begin{equation}\label{difeq2}
2Qf'(Q)-f(Q)=\{1-\frac{c^{2}\kappa^{2}}{m^{2}}(2\alpha-m\beta+m^{2}\gamma)\}Q
\end{equation}
The general solution of equation (\ref{difeq2}) is
\begin{equation}\label{Sol3}
f(Q)=c_{2}\sqrt{Q}+\{1-\frac{c^{2}\kappa^{2}}{m^{2}}(2\alpha-m\beta+m^{2}\gamma)\}Q
\end{equation}
where $c_{2}$ is an arbitrary constant.\\

\subsection{Case II} Here we consider the following scalar factor to introduce future singularity for the evolution of the universe \cite{Pasqua1, Rangdee, Odintsov1} which is
\begin{equation}\label{scale2}
a(t)=a_{0}(t_{s}-t)^{-n}
\end{equation}
Here $n>0$ is a constant and $t<t_{s}$ where $t_{s}$ denotes the
future singularity at a finite time. Using equations
(\ref{rhode1}), (\ref{go2}), and (\ref{scale2}) we get a
differential equation for $f(Q)$ as
\begin{equation}\label{difeq3}
2Qf'(Q)-f(Q)=\{1-\frac{c^{2}\kappa^{2}}{n}(\alpha n+\beta)\}Q
\end{equation}
The general solution of equation (\ref{difeq3}) for a non-flat
universe is given by,
\begin{equation}\label{Sol4}
f(Q)=c_{3}\sqrt{Q}+\{1-\frac{c^{2}\kappa^{2}}{n}(0.8824n+0.5016)\}Q
\end{equation}
and for a flat universe can be given by,
\begin{equation}\label{Sol5}
f(Q)=c_{3}\sqrt{Q}+\{1-\frac{c^{2}\kappa^{2}}{n}(0.8502n+0.4817)\}Q
\end{equation}
where $c_{3}$ is an arbitrary constant. Solving equations
(\ref{rhode1}), (\ref{cg1}), and (\ref{scale2}) we get a
differential equation of $f(Q)$ for the Chen-Jing model as
\begin{equation}\label{difeq4}
2Qf'(Q)-f(Q)=\{1-\frac{c^{2}\kappa^{2}}{n^{2}}(2\alpha+\beta
n+\gamma n^{2})\}Q
\end{equation}
The general solution of equation (\ref{difeq4}) is
\begin{equation}\label{Sol6}
f(Q)=c_{4}\sqrt{Q}+\{1-\frac{c^{2}\kappa^{2}}{n^{2}}(2\alpha+\beta
n+\gamma n^{2})\}Q
\end{equation}
where $c_{4}$ is an arbitrary constant.\\

\subsection{Case III} The analysis of intermediate inflation has become an interesting key to studying the analytic solution of corresponding potentials. According to \cite{Vallinotto, Khatua}, the scale factor for the case of intermediate inflation has been taken as
\begin{equation}\label{scale3}
a(t)=e^{Bt^{\theta}},
\end{equation}
where $0<\theta<1$ and $B>0$ are constants. Using equations
(\ref{rhode1}), (\ref{go2}) and (\ref{scale3}) we get a
differential equation of $f(Q)$ as
\begin{equation}\label{difeq5}
2Qf'(Q)-f(Q)=\{1-c^{2}\kappa^{2}(\alpha-\beta)\}Q-c^{2}\kappa^{2}\beta(B\theta)^{\frac{1}{\theta-1}}(\theta-1)\Big(\frac{Q}{6}\Big)^{\frac{\theta-2}{2(\theta-1)}}-c^{2}\kappa^{2}\beta
B\sqrt{\frac{Q}{6}}
\end{equation}
The general solution of equation (\ref{difeq5}) for a non-flat
universe is given by,
\begin{equation}\label{Sol7}
f(Q)=c_{5}\sqrt{Q}+(1-0.3808c^{2}\kappa^{2})Q+0.5016
c^{2}\kappa^{2}(B\theta)^{\frac{1}{\theta-1}}(\theta-1)^{2}\Big(\frac{Q}{6}\Big)^{\frac{\theta-2}{2(\theta-1)}}-0.5016
c^{2}\kappa^{2} B\sqrt{\frac{Q}{6}}\ln{Q}
\end{equation}
and for a flat universe is given by,
\begin{equation}\label{Sol8}
f(Q)=c_{5}\sqrt{Q}+(1-0.3685c^{2}\kappa^{2})Q+0.4817
c^{2}\kappa^{2}(B\theta)^{\frac{1}{\theta-1}}(\theta-1)^{2}\Big(\frac{Q}{6}\Big)^{\frac{\theta-2}{2(\theta-1)}}-0.4817
c^{2}\kappa^{2} B\sqrt{\frac{Q}{6}}\ln{Q}
\end{equation}
Using equations (\ref{rhode1}), (\ref{cg1}) and (\ref{scale3}) we
get a differential equation of $f(Q)$ for the Chen-Jing model as
\begin{equation*}
2Qf'(Q)-f(Q)=\{1-c^{2}\kappa^{2}(\gamma+2)\}Q-6c^{2}\kappa^{2}\alpha(B\theta)^{\frac{2}{\theta-1}}(\theta-1)(\theta-2)\sqrt{\frac{6}{Q}}+12\alpha
c^{2}\kappa^{2}\Big(\frac{Q}{6}\Big)^{\frac{1}{\theta-1}}
\end{equation*}
\begin{equation}\label{difeq6}
-6\beta
c^{2}\kappa^{2}(B\theta)^{\frac{1}{\theta-1}}(\theta-1)\Big(\frac{Q}{6}\Big)^{\frac{\theta-2}{2(\theta-1)}}
\end{equation}
The general solution of equation (\ref{difeq6}) is obtained as
\begin{equation*}
f(Q)=c_{6}\sqrt{Q}+\{1-c^{2}\kappa^{2}(\gamma+2)\}Q-3c^{2}\kappa^{2}\alpha(B\theta)^{\frac{2}{\theta-1}}(\theta-1)(\theta-2)\sqrt{\frac{6}{Q}}+12\alpha
c^{2}\kappa^{2}\frac{\theta-1}{3-\theta}\Big(\frac{Q}{6}\Big)^{\frac{1}{\theta-1}}
\end{equation*}
\begin{equation}\label{Sol9}
+6\beta
c^{2}\kappa^{2}(B\theta)^{\frac{1}{\theta-1}}(\theta-1)^{2}\Big(\frac{Q}{6}\Big)^{\frac{\theta-2}{2(\theta-1)}}
\end{equation}
which is the corresponding reconstructed $f(Q)$ model.

\subsection{Case IV}
To avoid the Big Bang singularity \cite{Rudra1, Gangopadhyay}, we
propose an embryonic universe model for the evolution as
\begin{equation}\label{scale4}
a(t)=A(B+e^{nt})^{\lambda}
\end{equation}
where $A$, $B$, $n$ and $\lambda$ are all model parameters. $A>0,
B>0$ is taken to make a positive scale factor and overcome all
singularities. If we keep $a<0$ or $\lambda<0$ then all
singularities remain. For having expanding model, we need to
consider $a>0$ and $\lambda>0$. Granda-Oliveros model is propelled
to investigate $f(Q)$ gravity for non-flat and flat universe.
Using equations (\ref{rhode1}), (\ref{go2}) and (\ref{scale4}) we
get a differential equation of $f(Q)$ as,
\begin{equation}\label{difeq7}
2Qf'(Q)-f(Q)=\{1-c^{2}\kappa^{2}(\alpha-\frac{\beta}{\lambda})\}Q-c^{2}\kappa^{2}\beta
n\sqrt{6Q}
\end{equation}
The general solution of equation (\ref{difeq7}) for a non-flat
universe is given by,
\begin{equation}\label{Sol10}
f(Q)=c_{7}\sqrt{Q}+\{1-c^{2}\kappa^{2}(0.8824-\frac{0.5016}{\lambda})\}Q+0.2508c^{2}\kappa^{2}n
\sqrt{6Q}\ln Q
\end{equation}
and for a flat universe is given by,
\begin{equation}\label{Sol11}
f(Q)=c_{7}\sqrt{Q}+\{1-c^{2}\kappa^{2}(0.8502-\frac{0.4817}{\lambda})\}Q+0.24085c^{2}\kappa^{2}n
\sqrt{6Q}\ln Q
\end{equation}
Using equations (\ref{rhode1}), (\ref{cg1}) and (\ref{scale4}) we
get a differential equation of $f(Q)$ for the Chen-Jing model as
\begin{equation}\label{difeq8}
2Qf'(Q)-f(Q)=\{1-c^{2}\kappa^{2}(\frac{4\alpha}{\lambda}-\frac{2\alpha}{\lambda^{2}}-\frac{\beta}{\lambda}+\gamma)\}Q+18n^{2}\alpha
\kappa^{2}c^{2}-\kappa^{2}c^{2}n\sqrt{6Q}(\frac{5\alpha}{\lambda}-4\alpha+\beta)
\end{equation}
The general solution of equation (\ref{difeq8}) is given by,
\begin{equation}\label{Sol12}
f(Q)=c_{8}\sqrt{Q}+\{1-c^{2}\kappa^{2}(\frac{4\alpha}{\lambda}-\frac{2\alpha}{\lambda^{2}}-\frac{\beta}{\lambda}+\gamma)\}Q-18\alpha
n^{2}\kappa^{2}c^{2}-\frac{\kappa^{2}c^{2}n}{2}\sqrt{6Q}(\frac{5\alpha}{\lambda}-4\alpha+\beta)\ln
Q.
\end{equation}
which is the reconstructed model.

\subsection{Case V}
In the quantum deformation of conformal group, $q$-de Sitter
deformation (quantum deformation of de Sitter universe) is
obtained in \cite{Lowe} to investigate dS/CFT correspondence. This
deformation has also been analyzed for the entrapment of entropy
and cosmology \cite{Lowe, Seo}. We consider this $q$-de Sitter
scalar factor for the evolution as \cite{Seo}
\begin{equation}\label{scale5}
a(t)=e_{q}(H_{0}t)=[1+(q-1)H_{0}t]^{\frac{1}{q-1}}
\end{equation}
This scale factor can interpolate the power law model as well as de Sitter spacetime model.\\
For early universe, $H_{0}t\gg1$ implies
\begin{equation}
a_{early}(t)\sim [H_{0}t]^{\frac{1}{q-1}}=t^{p}.
\end{equation}
Again for current accelerating universe, we have $p>1$ and $q<2$,
which corresponds to
\begin{equation}
a_{early}(t)\preceq e_{q}(H_{0}t)\preceq a_{dS}(t).
\end{equation}
By these two inequalities, $q$-de Sitter can associate the early universe with the present universe which plays an interesting cosmological scenario for this model.\\
Using equations (\ref{rhode1}), (\ref{go2}) and (\ref{scale5}) we
get a differential equation for $f(Q)$ as
\begin{equation}\label{difeq9}
2Qf'(Q)-f(Q)=[1-c^{2}\kappa^{2}\{\alpha-\beta(q-1)\}]Q
\end{equation}
The general solution of equation (\ref{difeq9}) for a non-flat
universe is given by,
\begin{equation}\label{Sol13}
f(Q)=c_{9}\sqrt{Q}+[1-c^{2}\kappa^{2}\{0.8824-0.5016(q-1)\}]Q
\end{equation}
and for a flat universe is given by,
\begin{equation}\label{Sol14}
f(Q)=c_{9}\sqrt{Q}+[1-c^{2}\kappa^{2}\{0.8502-0.4817(q-1)\}]Q
\end{equation}
Solving equations (\ref{rhode1}), (\ref{cg1}) and (\ref{scale5})
we get a differential equation of $f(Q)$ for the Chen-Jing model
as
\begin{equation}\label{difeq10}
2Qf'(Q)-f(Q)=[1-c^{2}\kappa^{2}\{\gamma-\beta(q-1)\}]Q-2c^{2}\kappa^{2}\alpha
H_{0}(q-1)^{2}\sqrt{6Q}
\end{equation}
The general solution of equation (\ref{difeq10}) is given by,
\begin{equation}\label{Sol15}
f(Q)=c_{10}\sqrt{Q}+[1-c^{2}\kappa^{2}\{\gamma-\beta(q-1)\}]Q-c^{2}\kappa^{2}\alpha
H_{0}(q-1)^{2}\sqrt{6Q}\ln Q.
\end{equation}
which is the reconstructed $f(Q)$ model for Chen-Jing dark energy.

\section{Energy Conditions}
The positiveness of the energy-momentum tensor in the presence of
matter can be determined using energy conditions. These
requirements not only address the causal and geodesic structure of
spacetime \cite{ec1}, but they also characterize the
attractiveness of gravity. Because the energy conditions lead to
several strong singularity theorems \cite{ec2}, it is recognized
that they are intimately related to General Relativity. Now, the
definition and formulation of energy requirements in the context
of modified gravity theories is very complex, with ramifications
that differ greatly from those found in general relativity. In
particular, the non-standard (fictitious) fluids associated with
the extra degrees of freedom of modified gravity should yield
intriguing results when combined with the energy conditions, which
provides us with some insights into gravity's non-attractive
nature and cosmic acceleration. This is significant because, as of
yet, we lack a cosmological model that is both free of all
cosmological problems and consistent with observations. The main
conclusions are that matter may exhibit additional thermodynamic
properties and that gravity may continue to be attractive even in
the presence of high negative pressures. However, for normal
matter, we can have repulsive gravity. We can formulate consistent
energy conditions for a wide range of theories because additional
degrees of freedom related to modified gravity theories fall under
the category of effective fluids. These are significant factors
from a cosmological perspective. As an illustration, we observe
that the existence of dark energy might be regarded as a clear
transgression of GR's basic definition of energy conditions. A
broader approach to modified gravity theory, however, does not
violate this; rather, it merely reinterprets the extra degrees of
freedom that result from the dynamics of the theory \cite{ec3}.
Thus, it is evident that researching energy conditions in changed
gravity has a lot to offer. Readers interested in further in-depth
discussions on energy conditions in modified gravity theories
might consult Refs.\cite{ec3, ec4}. Additionally, refs.\cite{ec5,
ec6, ec7, ec8, ec9} contain energy conditions in the background of
many modified gravity theories.

This section examines the energy requirements that the $f(Q)$ theory's thermodynamic parameters must meet, thus placing certain limitations on the model's parameters. It should be noted that, in order for the anti-gravitational effect to be effective, the late cosmic acceleration necessitates the violation of the strong energy condition. Dark energy could be the matter component causing this violation. The standard general theory of relativity can be used to determine the four main energy conditions. Considering isotropic cosmology they are obtained as:\\\\

(I) Weak Energy Condition (WEC)~$\Rightarrow$~ $\rho_{eff}\geq 0$,~~~~$\rho_{eff}+p_{eff}\geq 0$\\

(II) Null Energy Condition (NEC)~$\Rightarrow$~~~~~$\rho_{eff}+p_{eff}\geq 0$\\

(III) Dominant Energy condition (DEC)~$\Rightarrow$~ $\rho_{eff}\geq 0$,~~~~$\rho_{eff}\geq~ \mid p_{eff} \mid$\\

(IV) Strong Energy condition (SEC)~$\Rightarrow$~ ~~$\rho_{eff}+3p_{eff}\geq 0$\\

%\begin{equation}\label{den}
%\rho=\frac{f}{2}-6H^{2}f'(Q),
%\end{equation}

%\begin{equation}\label{pre}
%p=\Big(\dot{H}+3H^{2}+\frac{\dot{f'}(Q)}{f'(Q)}H\Big)2f'(Q)-%\frac{f}{2}.
%\end{equation}

%\begin{equation}
%3H^{2}=-\frac{1}{2}\rho_{eff}
%\end{equation}

%\begin{equation}
%3H^{2}+\dot{H}=\frac{p_{eff}}{2}
%\end{equation}

Using the Friedmann equations (\ref{frw1}), (\ref{frw2}),
(\ref{rhode1}) and (\ref{pde1}) we obtain the effective energy
density and pressure of the model as follows,

\begin{equation}\label{eff1}
\rho_{eff}=\frac{1}{f'(Q)}\Big(\rho+\frac{f}{2}\Big),
\end{equation}

\begin{equation}\label{eff2}
p_{eff}=-2\frac{\dot{f'}(Q)}{f'(Q)}H+\frac{1}{f'(Q)}\Big(p+\frac{f}{2}\Big).
\end{equation}
We will use the above expressions for energy density and pressure
to obtain the energy conditions for various cases discussed in the
paper. The main objective behind this is to validate our
reconstructed models against the attractive nature of gravity.

\subsection{Case I:}
Here we will explore the energy conditions for the first case. Using the equations (\ref{Sol1}), (\ref{Sol2}), (\ref{Sol3}), (\ref{eff1}) and (\ref{eff2}), we get the following inequalities for WEC, NEC, DEC, SEC for the GO model:\\

\begin{itemize}

\item \textbf{WEC, NEC and DEC:}

\textbf{WEC and DEC}

\begin{equation}
\frac{c_{1}(\sqrt{Q}-\frac{1}{2\sqrt{Q})}}{1+\frac{c_{1}}{2\sqrt{Q}}-\frac{c^{2}\kappa^{2}(\alpha
m-\beta)}{m}}\geq 0,
\end{equation}
and\\

\textbf{WEC and NEC}

\begin{equation}
-\frac{Q[c_{1}(m-6m^{2})+2(3m-1)\sqrt{Q}\{-c^{2}\beta\kappa^{2}+m(c^{2}\alpha\kappa^{2}-1)\}]}{3m\{c_{1}m+2\sqrt{Q}(m-c^{2}m\alpha\kappa^{2}+c^{2}\beta\kappa^{2}\}}\geq
0.
\end{equation}

\item \textbf{DEC:}

\begin{equation}
-\frac{c_{1}mQ}{c_{1}m+2\sqrt{Q}(m-c^{2}m\alpha\kappa^{2}+c^{2}\beta\kappa^{2})}+\left|Q-\frac{Q}{3m}\right|\leq
0,
\end{equation}

\item \textbf{SEC:}

\begin{equation}
    -\frac{Q[c_{1}(m-4m^{2})+2(-1+3m)\sqrt{Q}\{-c^{2}\beta\kappa^{2}+m(-1+c^{2}\alpha\kappa^{2})\}]}{m\{c_{1}m+2\sqrt{Q}(m-c^{2}m\alpha\kappa^{2}+c^{2}\beta\kappa^{2})\}}\geq 0.
\end{equation}
\end{itemize}

For the Chen-Jing model the corresponding expressions read as,

\begin{itemize}
    \item \textbf{WEC, NEC and DEC:}

    \textbf{WEC and DEC}

    \begin{equation}
\frac{c_{2}\left(\sqrt{Q}-\frac{1}{2\sqrt{Q}}\right)}{1+\frac{c_{2}}{2\sqrt{Q}}-\frac{c^{2}\kappa^{2}(2\alpha-m\beta+m^{2}\gamma)}{m^{2}}}\geq
0,
\end{equation}
and\\
\textbf{WEC and NEC}
\begin{equation}
-\frac{Q[c_{2}\sqrt{Q}(m^{2}-6m^{3})+2(3m-1)\sqrt{Q}\{2c^{2}\alpha\kappa^{2}-c^{2}m\beta\kappa^{2}+m^{2}(c^{2}\gamma\kappa^{2}-1)\}]}{3m[c_{2}m^{2}+2\sqrt{Q}\{-2c^{2}\alpha\kappa^{2}+c^{2}m\beta\kappa^{2}+m^{2}(1-c^{2}\gamma\kappa^{2})\}]}\geq
0.
\end{equation}

\item \textbf{DEC:}

\begin{equation}
    -\frac{c_{2}\sqrt{Q}}{2\Big(1+\frac{c_{2}}{2\sqrt{Q}}-\frac{c^{2}\{2\alpha+m(m\gamma-\beta)\}\kappa^{2}}{m^{2}}\Big)}+\left|Q-\frac{Q}{3m}\right|\leq 0.
\end{equation}

\item \textbf{SEC:}

\begin{equation}
    -\frac{Q[c_{2}(m^{2}-4m^{3})+2(-1+3m)\sqrt{Q}\{2c^{2}\alpha\kappa^{2}-c^{2}m\beta\kappa^{2}+m^{2}(-1+c^{2}\gamma\kappa^{2})\}]}{m[c_{2}m^{2}+2\sqrt{Q}\{-2c^{2}\alpha\kappa^{2}+c^{2}m\beta\kappa^{2}+m^{2}(1-c^{2}\gamma\kappa^{2})\}]}\geq 0.
\end{equation}
\end{itemize}

The above energy conditions are plotted in Fig.(1) to get deeper insights into these expressions.\\\\

\begin{figure}
\includegraphics[height=2in,width=2in]{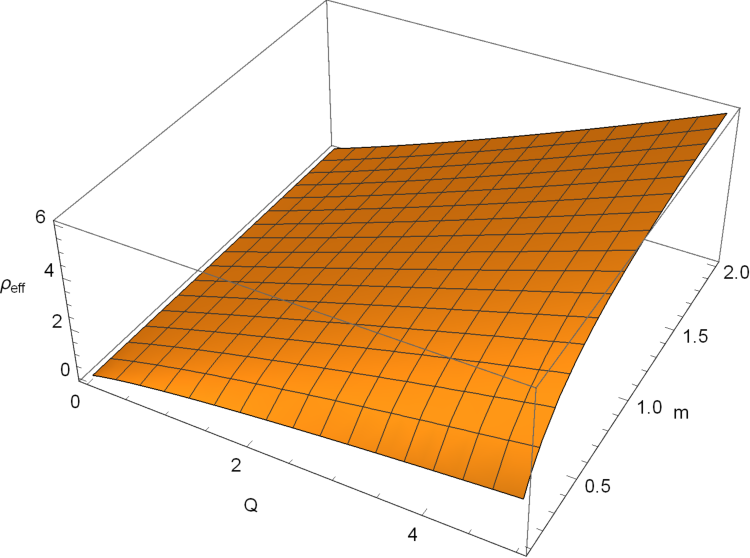}~~~~
\includegraphics[height=2in,width=2in]{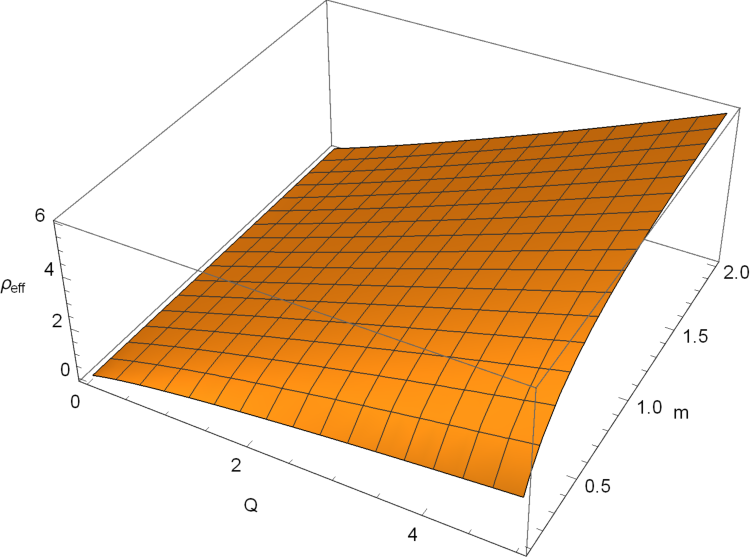}~~~~
\includegraphics[height=2in,width=2in]{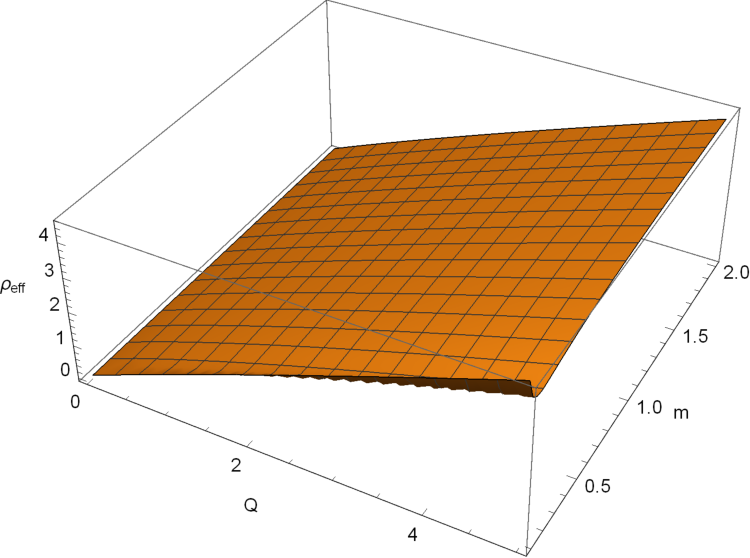}

~~~~~~~~~~~~~~~~~~(1a)~~~~~~~~~~~~~~~~~~~~~~~~~~~~~~~~~~~~~~~~~~~~~~~(1b)~~~~~~~~~~~~~~~~~~~~~~~~~~~~~~~~~~~~~~~~~~~~~~~(1c)\\

\vspace{1 mm}

\vspace{1 mm}
\end{figure}

\begin{figure}
\includegraphics[height=2in,width=2in]{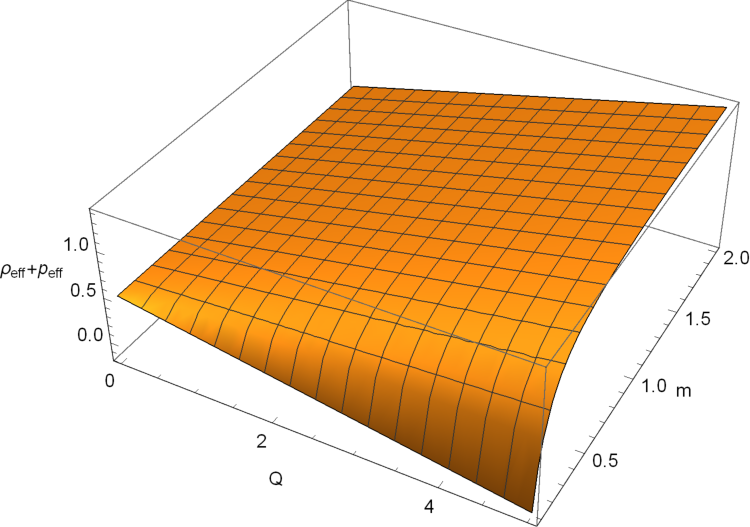}~~~~
\includegraphics[height=2in,width=2in]{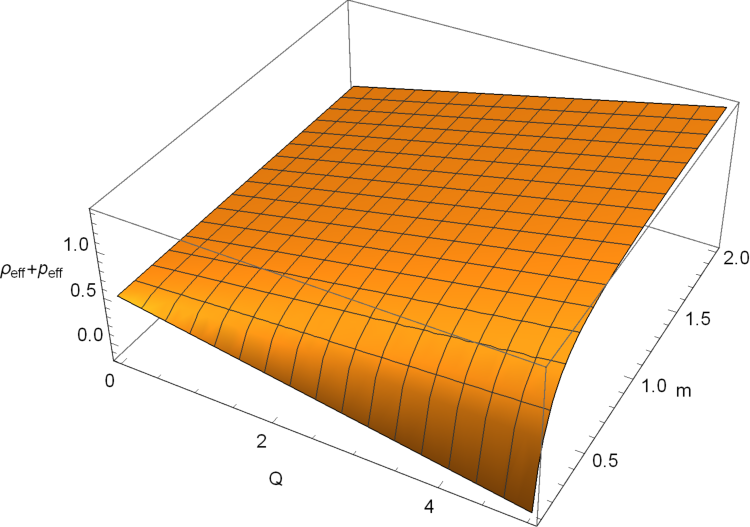}~~~~
\includegraphics[height=2in,width=2in]{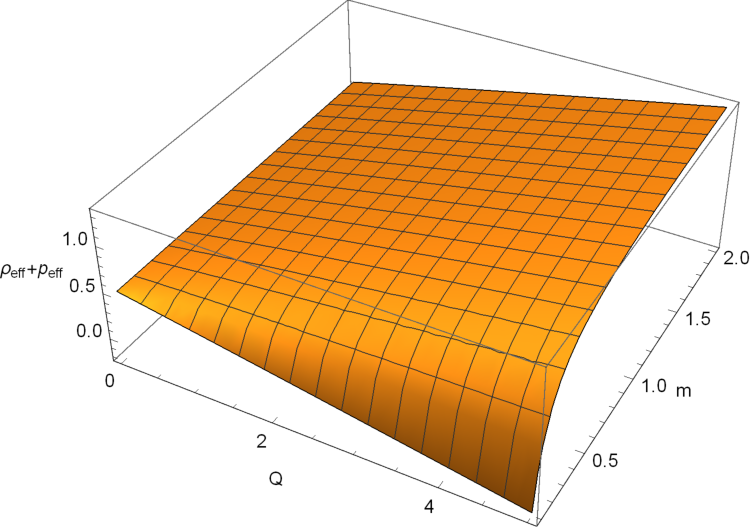}

~~~~~~~~~~~~~~~~~~(1d)~~~~~~~~~~~~~~~~~~~~~~~~~~~~~~~~~~~~~~~~~~~~~~~(1e)~~~~~~~~~~~~~~~~~~~~~~~~~~~~~~~~~~~~~~~~~~~~~~~(1f)\\

\vspace{1 mm}

\vspace{1 mm}
\end{figure}

\begin{figure}
\includegraphics[height=2in,width=2in]{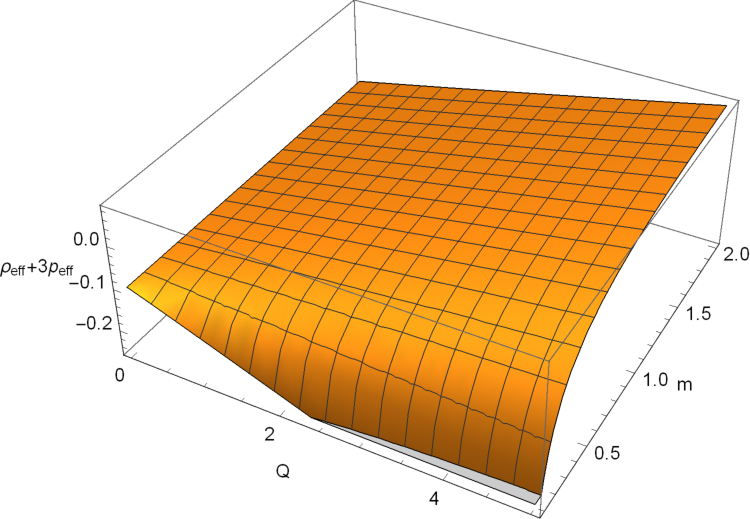}~~~~
\includegraphics[height=2in,width=2in]{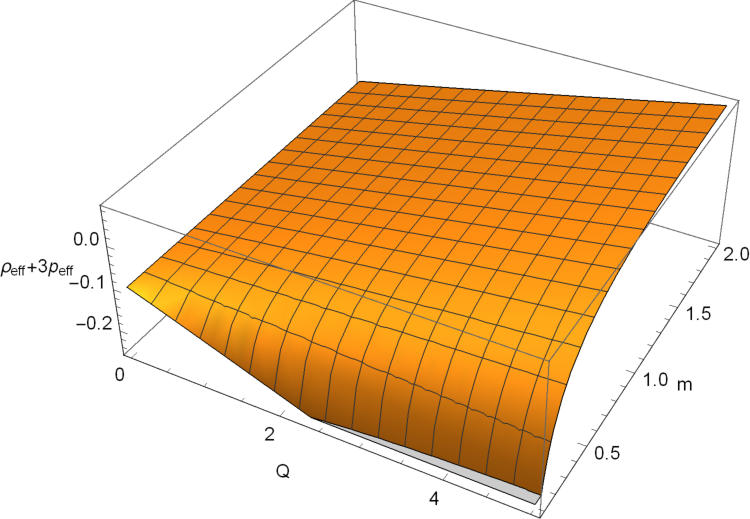}~~~~
\includegraphics[height=2in,width=2in]{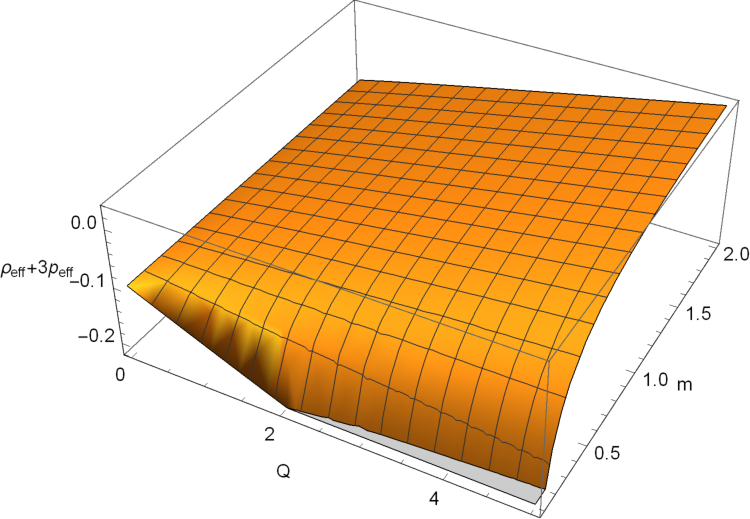}

~~~~~~~~~~~~~~~~~~(1g)~~~~~~~~~~~~~~~~~~~~~~~~~~~~~~~~~~~~~~~~~~~~~~~(1h)~~~~~~~~~~~~~~~~~~~~~~~~~~~~~~~~~~~~~~~~~~~~~~~(1i)\\

\vspace{1 mm}

\vspace{1 mm}
\end{figure}

\begin{figure}
\includegraphics[height=2in,width=2in]{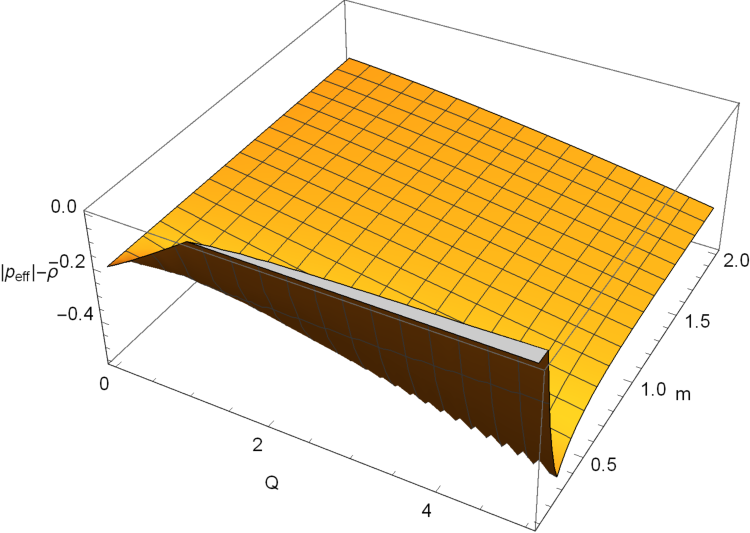}~~~~
\includegraphics[height=2in,width=2in]{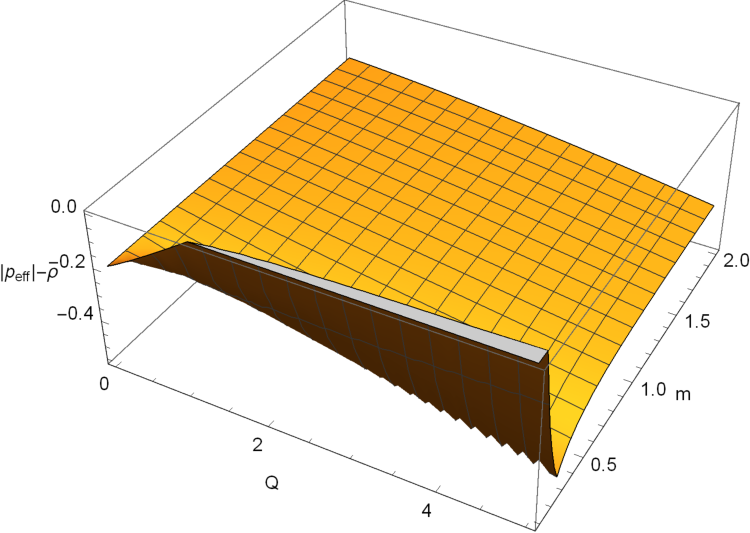}~~~~
\includegraphics[height=2in,width=2in]{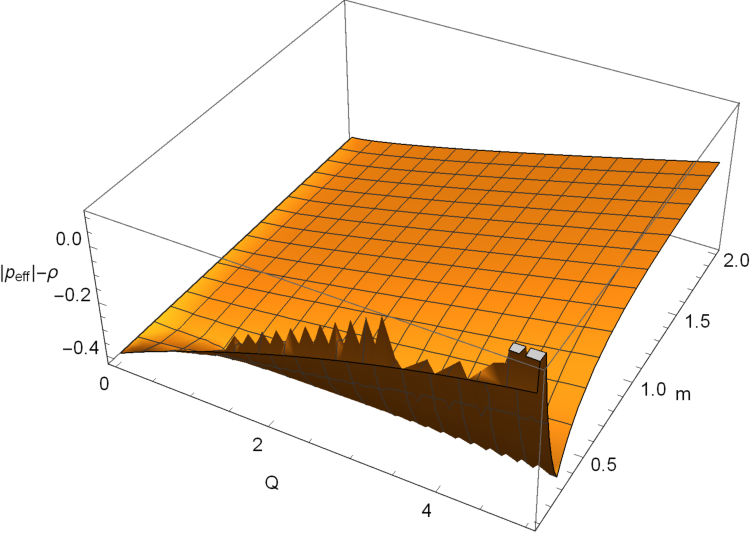}

~~~~~~~~~~~~~~~~~~(1j)~~~~~~~~~~~~~~~~~~~~~~~~~~~~~~~~~~~~~~~~~~~~~~~(1k)~~~~~~~~~~~~~~~~~~~~~~~~~~~~~~~~~~~~~~~~~~~~~~~(1l)\\

\vspace{1 mm}
\textbf{Fig.1:~(a)-(c)}: Scenario of WEC ($\rho_{eff}\geq 0$) for GO non-flat, flat and Chen-Jing models, \textbf{Fig.1:~(d)-(f)}: Scenario of NEC ($\rho_{eff}+p_{eff}\geq 0$) for GO non-flat, flat and Chen-Jing models, \textbf{Fig.1:~(g)-(i)}: Scenario of SEC ($\rho_{eff}+3p_{eff}\geq 0$) for GO non-flat, flat and Chen-Jing models and \textbf{Fig.1:~(j)-(l)}: Scenario of DEC ($|p_{eff}|-\rho_{eff}\leq 0$) for GO non-flat, flat and Chen-Jing models with the variation of $Q$ and $m$ for the different parameters like $c$, $c_{1}$, $c_{2}$, $\kappa=1$, $\alpha$, $\beta$ and $\gamma$ for \textbf{Case I}.\\
\vspace{2 mm}
\end{figure}

\subsection{Case II:}
Here we will study the energy conditions for the second case. Using the equations (\ref{Sol4}), (\ref{Sol5}), (\ref{Sol6}), (\ref{eff1}) and (\ref{eff2}), we get the following inequalities for WEC, NEC, DEC, SEC for the GO model:\\

\begin{itemize}

\item \textbf{WEC, NEC and DEC:}

\textbf{WEC and DEC}

\begin{equation}
  \frac{c_{3}nQ}{c_{3}+2\sqrt{Q}(n-c^{2}n\alpha\kappa^{2}+c^{2}\beta\kappa^{2})}\geq 0,
\end{equation}
and\\

\textbf{WEC and NEC}

\begin{equation}
    -\frac{Q[-c_{3}n(1+6n)+2(1+3n)\sqrt{Q}\{-c^{2}\beta\kappa^{2}+n(-1+c^{2}\alpha\kappa^{2})\}]}{3n\{c_{3}n+2\sqrt{Q}(n-c^{2}n\alpha\kappa^{2}+c^{2}\beta\kappa^{2}}\geq 0.
\end{equation}

\item \textbf{DEC:}

\begin{equation}
    -\frac{c_{3}nQ}{c_{3}n+2\sqrt{Q}(n-c^{2}n\alpha\kappa^{2}+c^{2}\beta\kappa^{2})}+\left|Q+\frac{Q}{3n}\right|\leq 0.
\end{equation}

\item \textbf{SEC:}
\begin{equation}
    -\frac{Q[(-c_{3}n(1+4n)+2(1+3n)\sqrt{Q}\{-c^{2}\beta\kappa^{2}+n(-1+c^{2}\alpha\kappa^{2})\}]}{n\{c_{3}n+2\sqrt{Q}(n-c^{2}n\alpha\kappa^{2}+c^{2}\beta\kappa^{2}\}}\geq 0.
\end{equation}

\end{itemize}

The corresponding expressions for the Chen-Jing model are obtained as,\\

\begin{itemize}
    \item \textbf{WEC, NEC and DEC:}

    \textbf{WEC and DEC}

    \begin{equation}
        \frac{c_{4}n^{2}Q}{c_{4}n^{2}-2\sqrt{Q}\{2c^{2}\alpha\kappa^{2}+c^{2}n\beta\kappa^{2}+n^{2}(-1+c^{2}\gamma\kappa^{2})\}}\geq 0,
    \end{equation}
and\\

\textbf{WEC and NEC}

\begin{equation}
    \frac{Q[-c_{4}n^{2}(1+6n)+2(1+3n)\sqrt{Q}\{2c^{2}\alpha\kappa^{2}+c^{2}n\beta\kappa^{2}+n^{2}(-1+c^{2}\gamma\kappa^{2})\}]}{3n[-c_{4}n^{2}+2\sqrt{Q}\{2c^{2}\alpha\kappa^{2}+c^{2}n\beta\kappa^{2}+n^{2}(-1+c^{2}\gamma\kappa^{2})\}]}\geq 0.
\end{equation}

\item \textbf{DEC:}

\begin{equation}
    -\frac{c_{4}\sqrt{Q}}{2\left[1+\frac{c_{4}}{2\sqrt{Q}}-\frac{c^{2}\{2\alpha+n(\beta+n\gamma)\}\kappa^{2}}{n^{2}}\right]}+\left|Q+\frac{Q}{3n}\right|\leq 0.
\end{equation}

\item \textbf{SEC:}
\begin{equation}
    \frac{Q\left[-c_{4}n^{2}(1+4n)+2(1+3n)\sqrt{Q}\{2c^{2}\alpha\kappa^{2}+c^{2}n\beta\kappa^{2}+n^{2}(-1+c^{2}\gamma\kappa^{2})\}\right]}{n[-c_{4}n^{2}+2\sqrt{Q}\{2c^{2}\alpha\kappa^{2}+c^{2}n\beta\kappa^{2}+n^{2}(-1+c^{2}\gamma\kappa^{2})\}]}\geq 0.
\end{equation}

\end{itemize}

The above energy conditions are plotted in Fig.(2) to get deeper insights into these expressions.\\\\

\begin{figure}
\includegraphics[height=2in,width=2in]{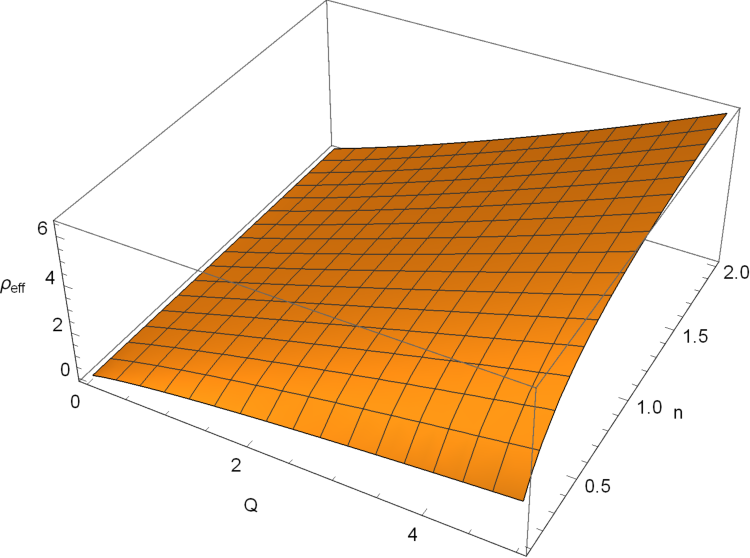}~~~~
\includegraphics[height=2in,width=2in]{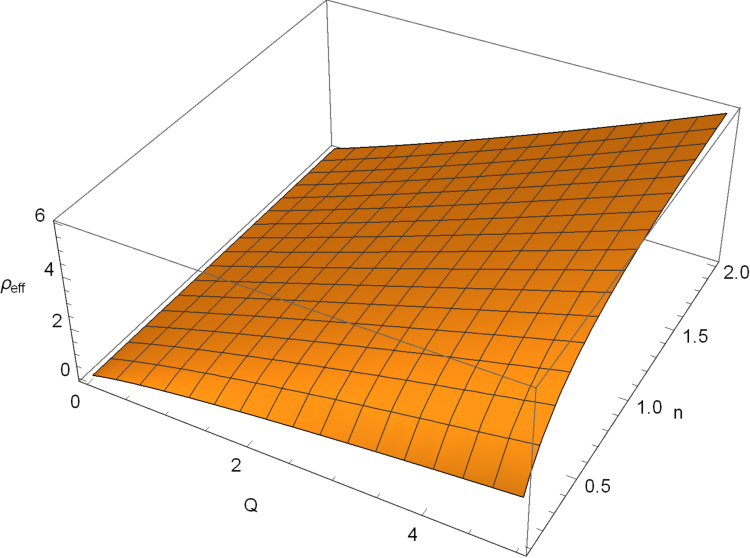}~~~~
\includegraphics[height=2in,width=2in]{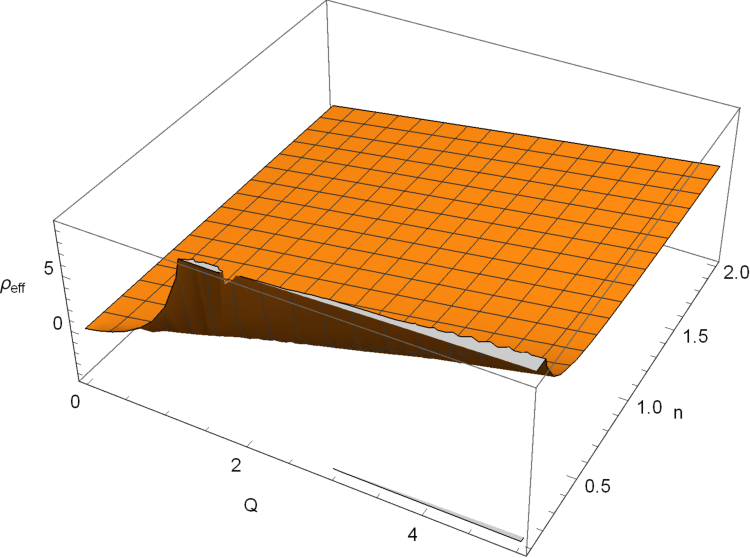}

~~~~~~~~~~~~~~~~~~(2a)~~~~~~~~~~~~~~~~~~~~~~~~~~~~~~~~~~~~~~~~~~~~~~~(2b)~~~~~~~~~~~~~~~~~~~~~~~~~~~~~~~~~~~~~~~~~~~~~~~(2c)\\

\vspace{1 mm}

\vspace{1 mm}
\end{figure}

\begin{figure}
\includegraphics[height=2in,width=2in]{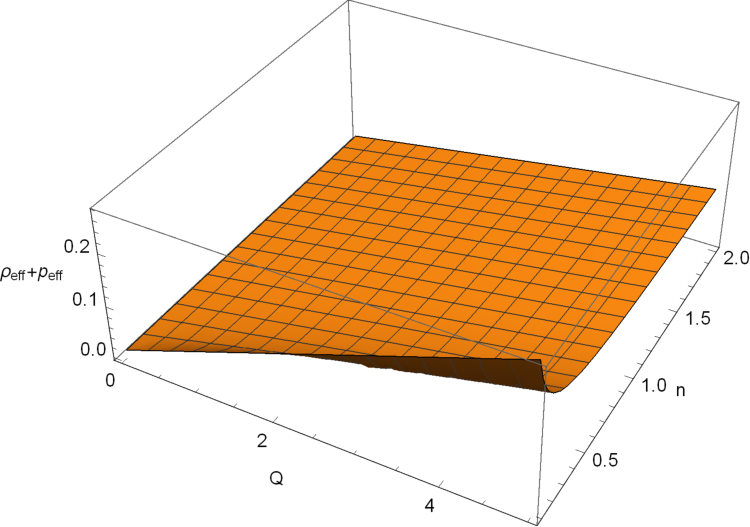}~~~~
\includegraphics[height=2in,width=2in]{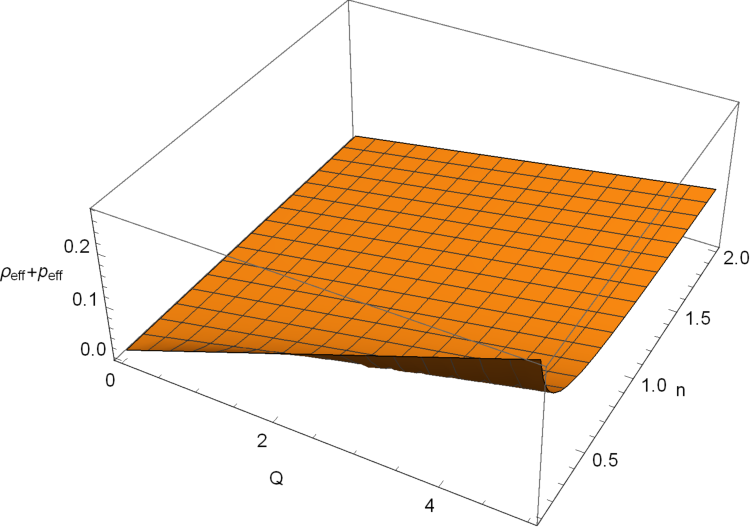}~~~~
\includegraphics[height=2in,width=2in]{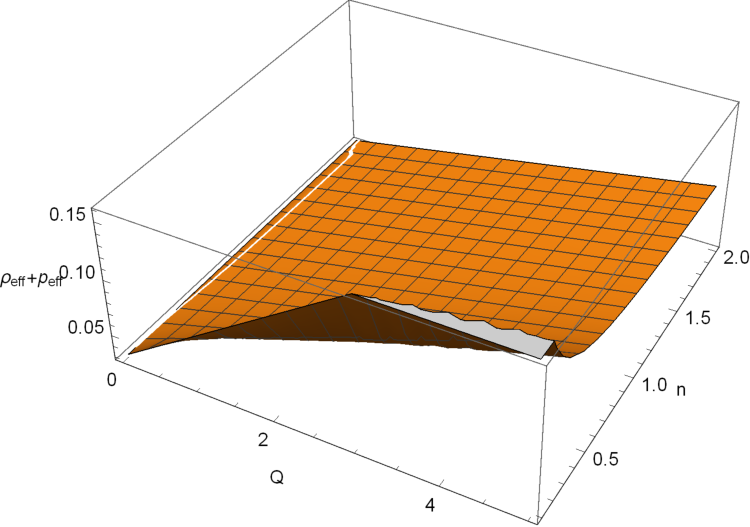}

~~~~~~~~~~~~~~~~~~(2d)~~~~~~~~~~~~~~~~~~~~~~~~~~~~~~~~~~~~~~~~~~~~~~~(2e)~~~~~~~~~~~~~~~~~~~~~~~~~~~~~~~~~~~~~~~~~~~~~~~(2f)\\

\vspace{1 mm}

\vspace{1 mm}
\end{figure}

\begin{figure}
\includegraphics[height=2in,width=2in]{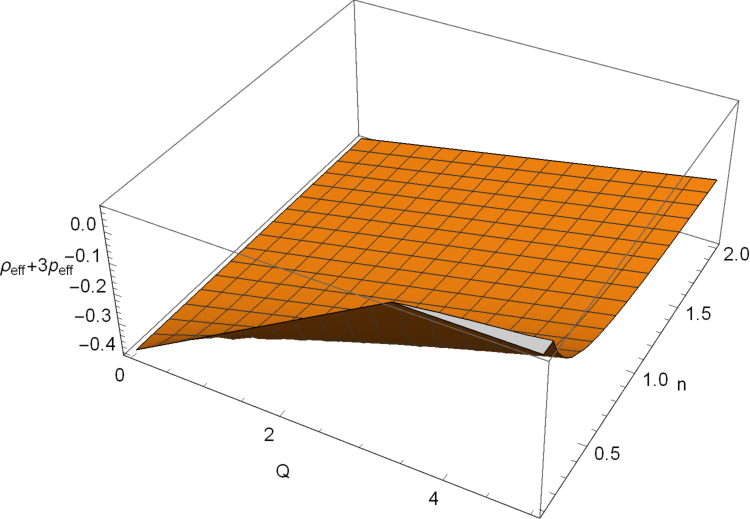}~~~~
\includegraphics[height=2in,width=2in]{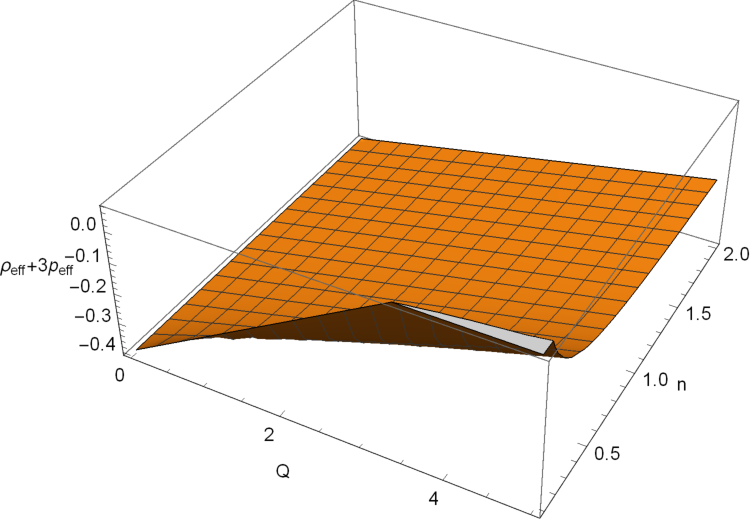}~~~~
\includegraphics[height=2in,width=2in]{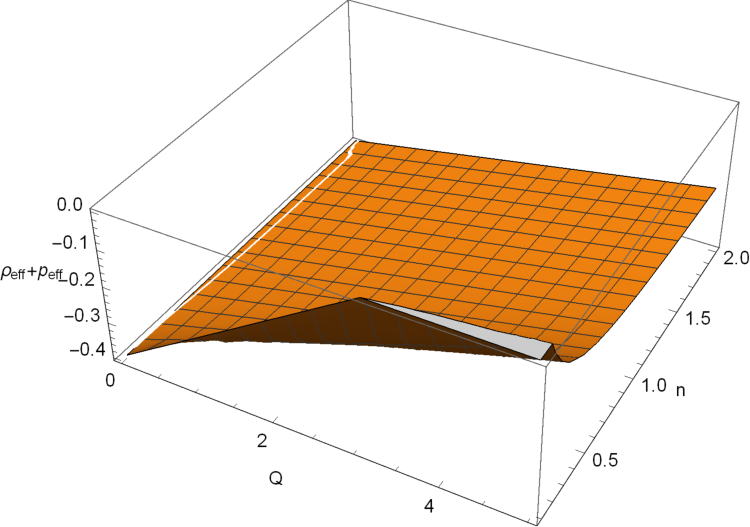}

~~~~~~~~~~~~~~~~~~(2g)~~~~~~~~~~~~~~~~~~~~~~~~~~~~~~~~~~~~~~~~~~~~~~~(2h)~~~~~~~~~~~~~~~~~~~~~~~~~~~~~~~~~~~~~~~~~~~~~~~(2i)\\

\vspace{1 mm}

\vspace{1 mm}
\end{figure}

\begin{figure}
\includegraphics[height=2in,width=2in]{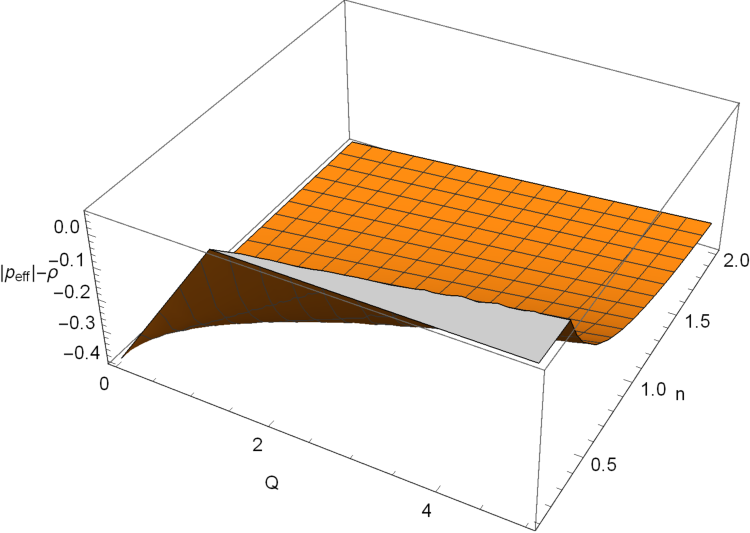}~~~~
\includegraphics[height=2in,width=2in]{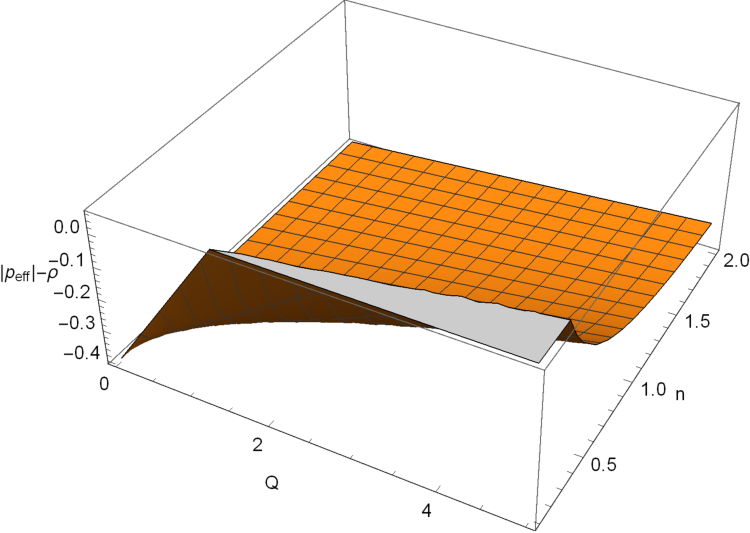}~~~~
\includegraphics[height=2in,width=2in]{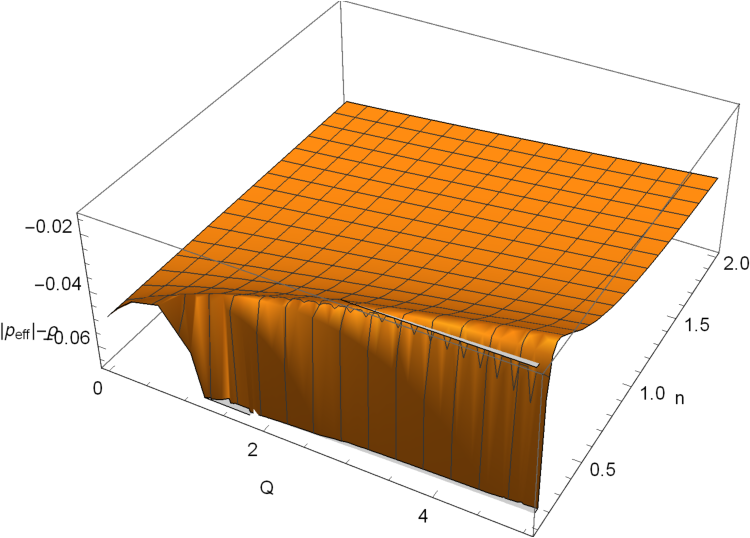}

~~~~~~~~~~~~~~~~~~(2j)~~~~~~~~~~~~~~~~~~~~~~~~~~~~~~~~~~~~~~~~~~~~~~~(2k)~~~~~~~~~~~~~~~~~~~~~~~~~~~~~~~~~~~~~~~~~~~~~~~(2l)\\

\vspace{1 mm}
\textbf{Fig.2:~(a)-(c)}: Scenario of WEC ($\rho_{eff}\geq 0$) for GO non-flat, flat and Chen-Jing models,
\textbf{Fig.2:~(d)-(f)}: Scenario of NULL ($\rho_{eff}+p_{eff}\geq 0$) for GO non-flat, flat and Chen-Jing models,
\textbf{Fig.2:~(g)-(i)}: Scenario of SEC ($\rho_{eff}+3p_{eff}\leq 0$) for GO non-flat, flat and Chen-Jing models and
\textbf{Fig.2:~(j)-(l)}: Scenario of DEC ($|p_{eff}|-\rho_{eff}\leq 0$) for GO non-flat, flat and Chen-Jing models
with the variation of $Q$ and $m$ for the different parameters like $c$, $c_{1}$, $c_{2}$, $\kappa=1$, $\alpha$, $\beta$ and $\gamma$ for \textbf{Case II}.\\
\vspace{2 mm}
\end{figure}

\subsection{Case III:}

In this section, we will investigate the energy conditions for the
third case. The expressions for the energy conditions obtained in
this case are very large. So in order to keep the paper in a
proper shape we have reported them in the \textbf{Appendix}. For
the comprehensive understanding of the physical aspects the energy
conditions are plotted in Fig.(3) to get deeper
insights into these expressions.\\\\

\begin{figure}
\includegraphics[height=2in,width=2in]{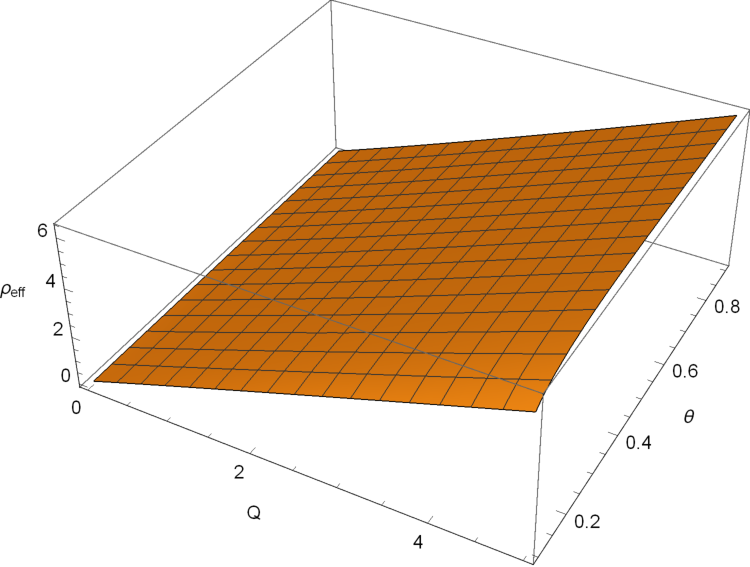}~~~~
\includegraphics[height=2in,width=2in]{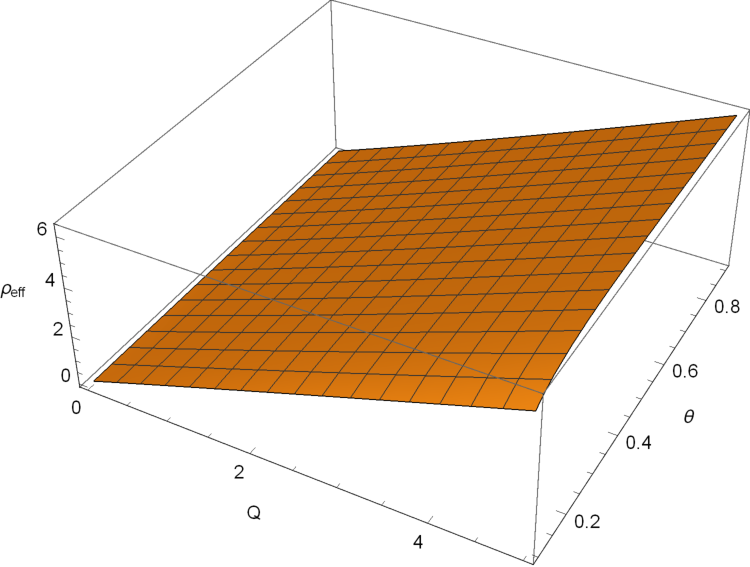}~~~~
\includegraphics[height=2in,width=2in]{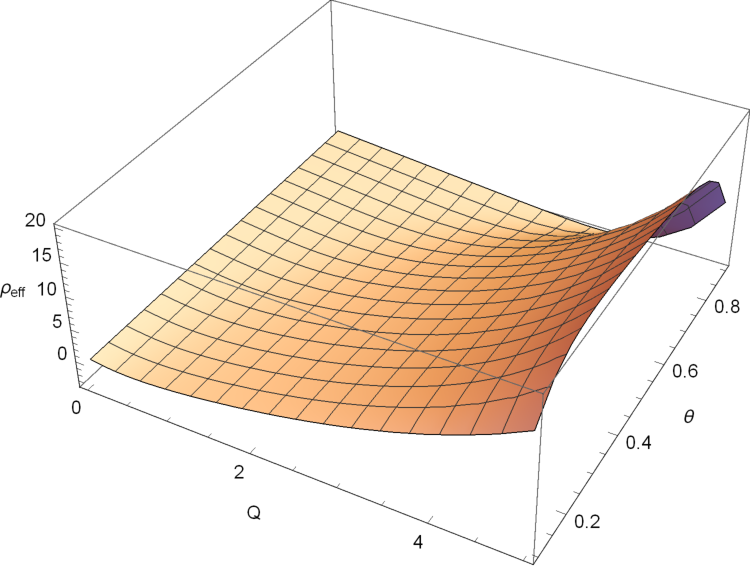}

~~~~~~~~~~~~~~~~~~(3a)~~~~~~~~~~~~~~~~~~~~~~~~~~~~~~~~~~~~~~~~~~~~~~~(3b)~~~~~~~~~~~~~~~~~~~~~~~~~~~~~~~~~~~~~~~~~~~~~~~(3c)\\

\vspace{1 mm}

\vspace{1 mm}
\end{figure}

\begin{figure}
\includegraphics[height=2in,width=2in]{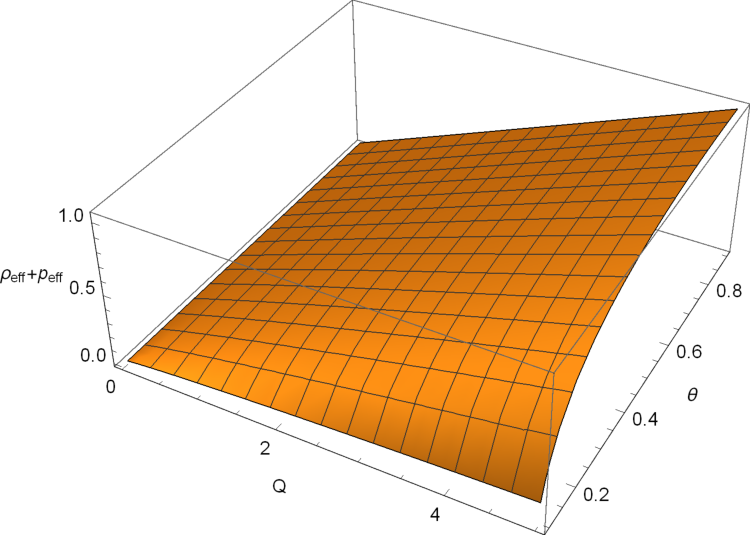}~~~~
\includegraphics[height=2in,width=2in]{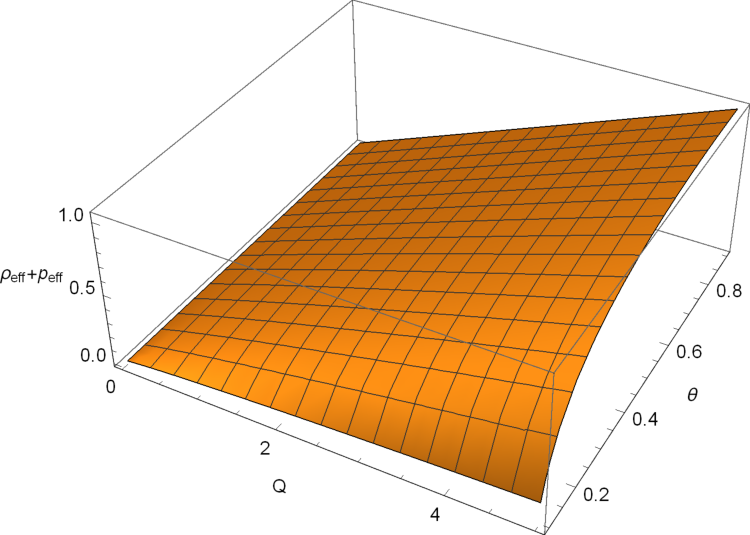}~~~~
\includegraphics[height=2in,width=2in]{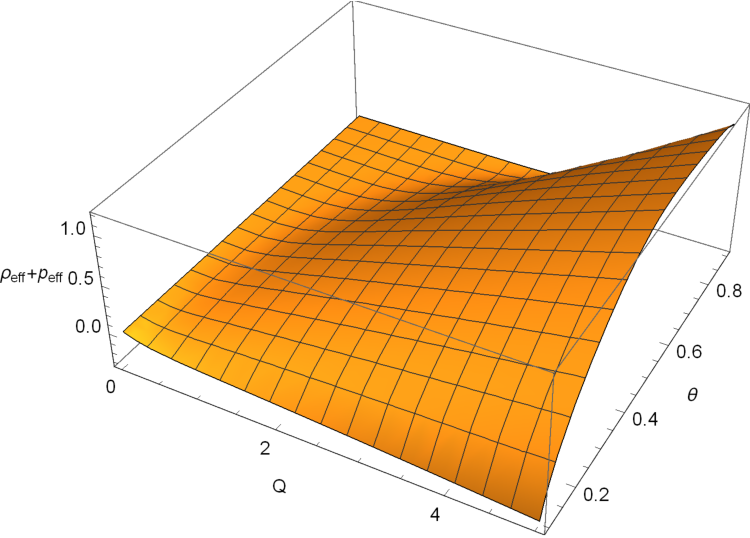}

~~~~~~~~~~~~~~~~~~(3d)~~~~~~~~~~~~~~~~~~~~~~~~~~~~~~~~~~~~~~~~~~~~~~~(3e)~~~~~~~~~~~~~~~~~~~~~~~~~~~~~~~~~~~~~~~~~~~~~~~(3f)\\

\vspace{1 mm}

\vspace{1 mm}
\end{figure}

\begin{figure}
\includegraphics[height=2in,width=2in]{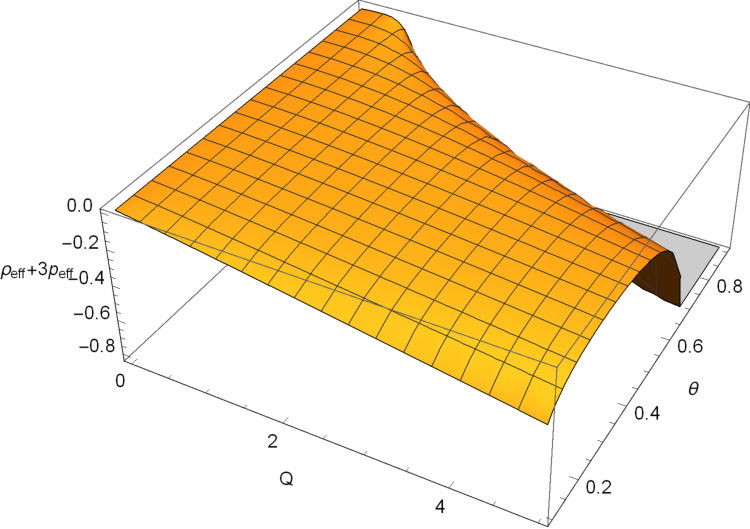}~~~~
\includegraphics[height=2in,width=2in]{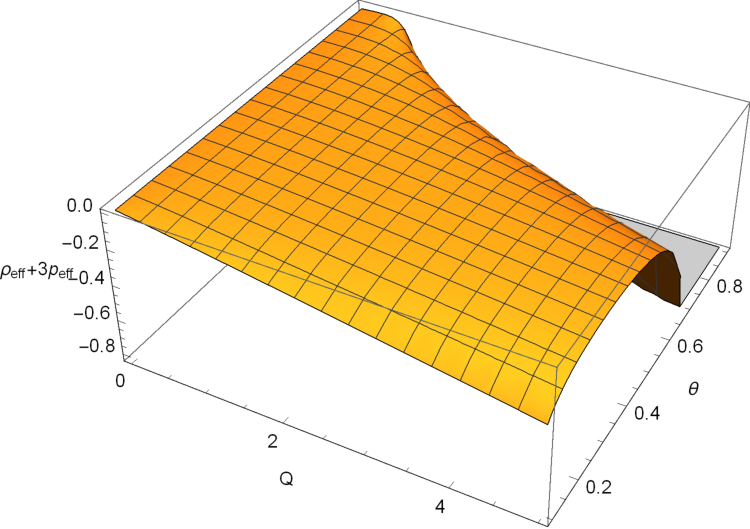}~~~~
\includegraphics[height=2in,width=2in]{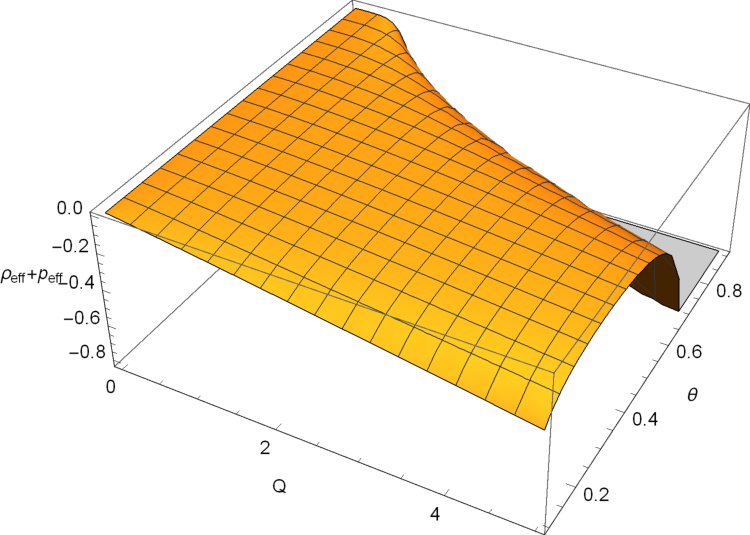}

~~~~~~~~~~~~~~~~~~(3g)~~~~~~~~~~~~~~~~~~~~~~~~~~~~~~~~~~~~~~~~~~~~~~~(3h)~~~~~~~~~~~~~~~~~~~~~~~~~~~~~~~~~~~~~~~~~~~~~~~(3i)\\

\vspace{1 mm}

\vspace{1 mm}
\end{figure}

\begin{figure}
\includegraphics[height=2in,width=2in]{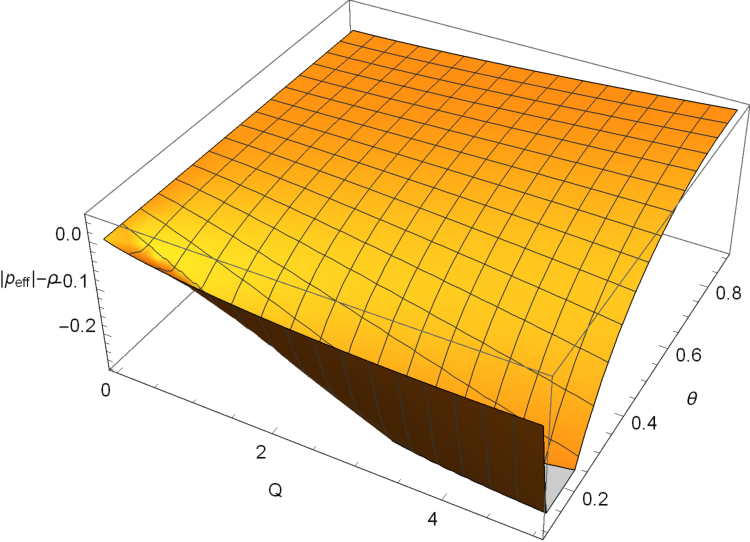}~~~~
\includegraphics[height=2in,width=2in]{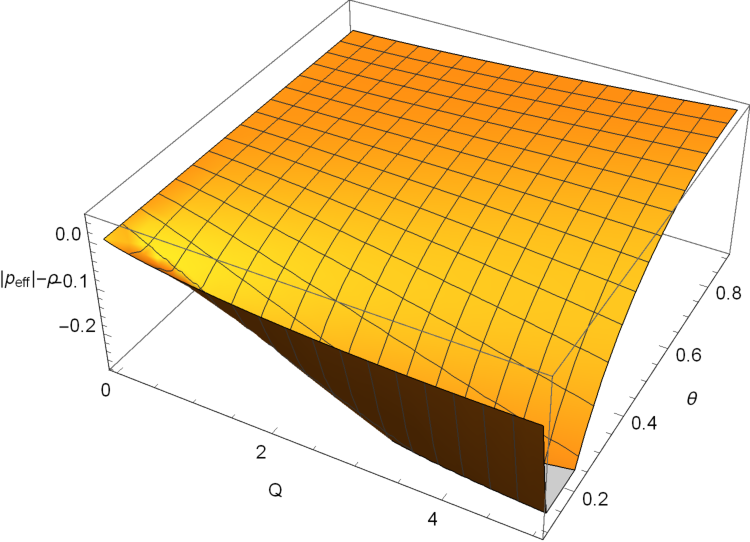}~~~~
\includegraphics[height=2in,width=2in]{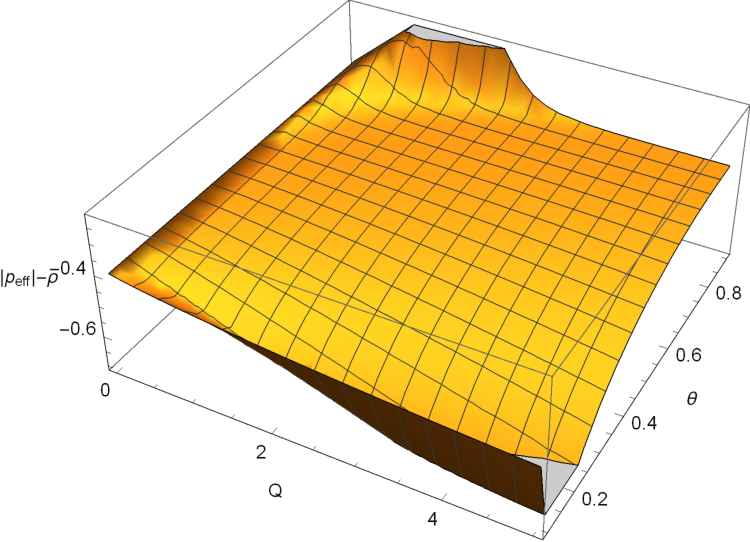}

~~~~~~~~~~~~~~~~~~(3j)~~~~~~~~~~~~~~~~~~~~~~~~~~~~~~~~~~~~~~~~~~~~~~~(3k)~~~~~~~~~~~~~~~~~~~~~~~~~~~~~~~~~~~~~~~~~~~~~~~(3l)\\

\vspace{1 mm}
\textbf{Fig.3:~(a)-(c)}: Scenario of WEC ($\rho_{eff}\geq 0$) for GO non-flat, flat and Chen-Jing models,
\textbf{Fig.3:~(d)-(f)}: Scenario of NULL ($\rho_{eff}+p_{eff}\geq 0$) for GO non-flat, flat and Chen-Jing models,
\textbf{Fig.3:~(g)-(i)}: Scenario of SEC ($\rho_{eff}+3p_{eff}\leq 0$) for GO non-flat, flat and Chen-Jing models and
\textbf{Fig.3:~(j)-(l)}: Scenario of DEC ($|p_{eff}|-\rho_{eff}\leq 0$) for GO non-flat, flat and Chen-Jing models with
the variation of $Q$ and $m$ for the different parameters like $c$, $c_{1}$, $c_{2}$, $\kappa=1$, $\alpha$, $\beta$ and $\gamma$ for \textbf{Case III}.\\
\vspace{2 mm}
\end{figure}

\subsection{Case IV:}
Here we will report the energy conditions for the fourth case.
Just like the previous case here also the expressions are very
large. Hence we have reported them in the \textbf{Appendix}. For a comprehensive understanding the energy conditions are
plotted in Fig.(4).\\\\

\begin{figure}
\includegraphics[height=2in,width=2in]{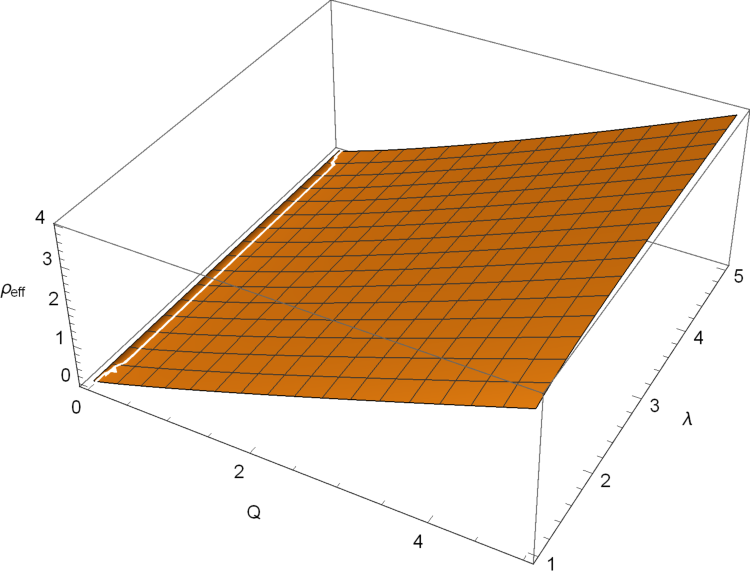}~~~~
\includegraphics[height=2in,width=2in]{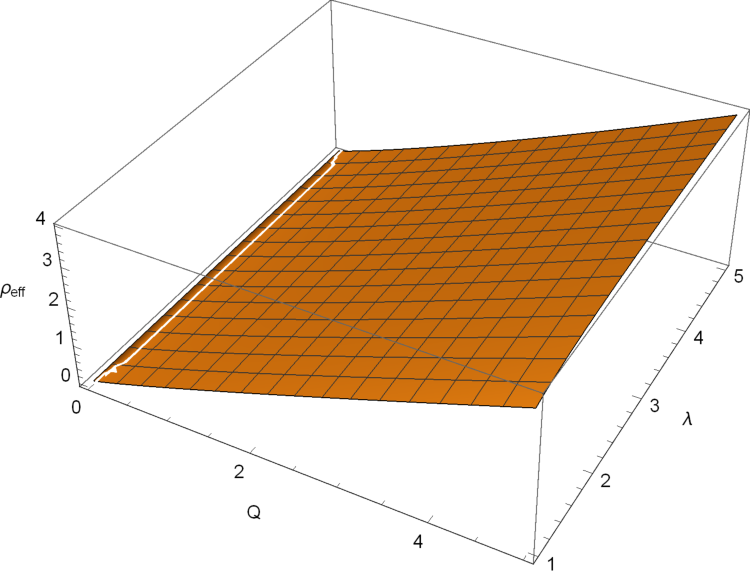}~~~~
\includegraphics[height=2in,width=2in]{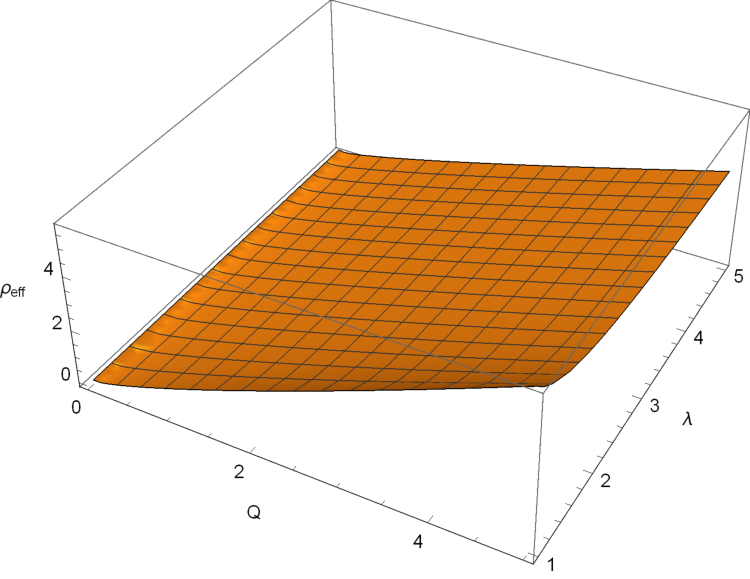}

~~~~~~~~~~~~~~~~~~(4a)~~~~~~~~~~~~~~~~~~~~~~~~~~~~~~~~~~~~~~~~~~~~~~~(4b)~~~~~~~~~~~~~~~~~~~~~~~~~~~~~~~~~~~~~~~~~~~~~~~(4c)\\

\vspace{1 mm}

\vspace{1 mm}
\end{figure}

\begin{figure}
\includegraphics[height=2in,width=2in]{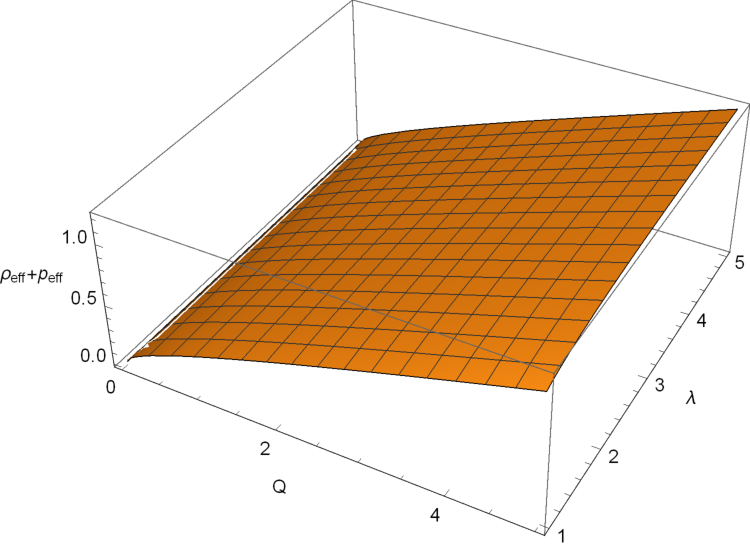}~~~~
\includegraphics[height=2in,width=2in]{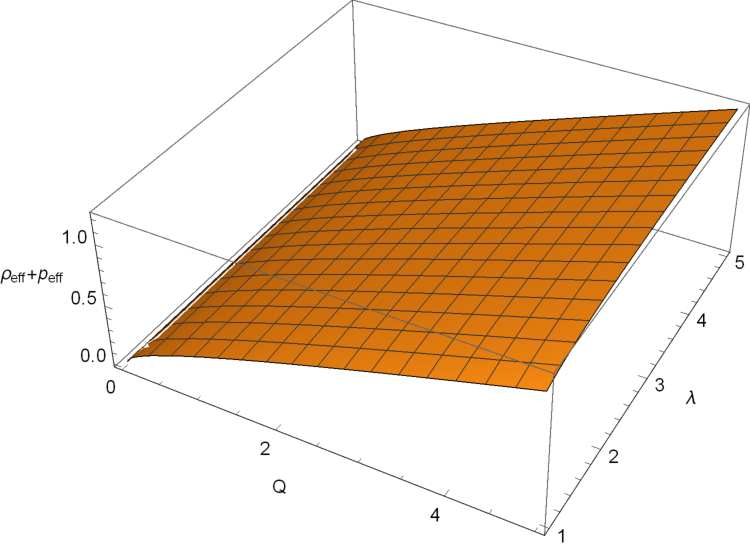}~~~~
\includegraphics[height=2in,width=2in]{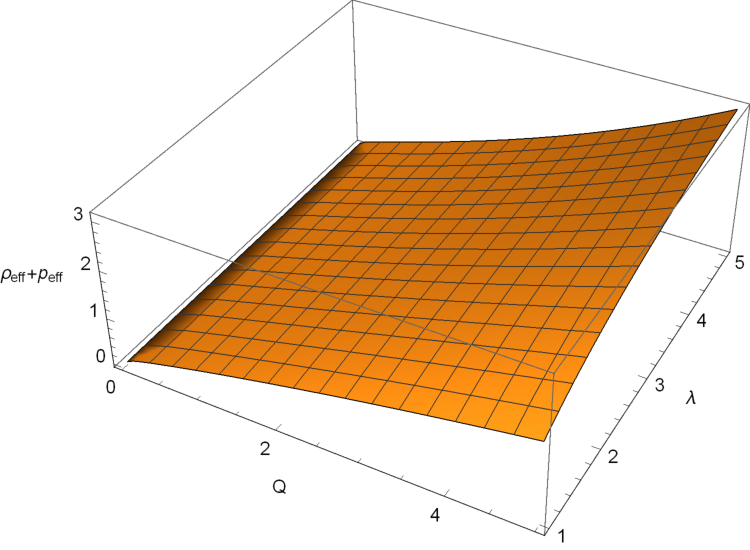}

~~~~~~~~~~~~~~~~~~(4d)~~~~~~~~~~~~~~~~~~~~~~~~~~~~~~~~~~~~~~~~~~~~~~~(4e)~~~~~~~~~~~~~~~~~~~~~~~~~~~~~~~~~~~~~~~~~~~~~~~(4f)\\

\vspace{1 mm}

\vspace{1 mm}
\end{figure}

\begin{figure}
\includegraphics[height=2in,width=2in]{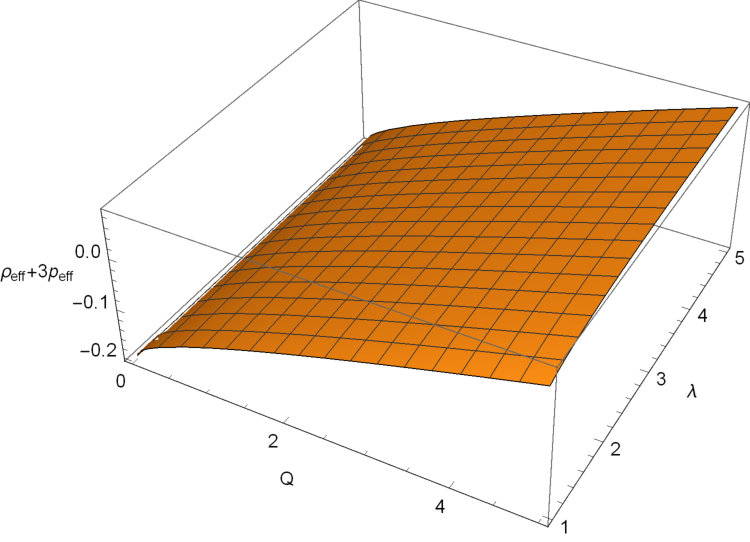}~~~~
\includegraphics[height=2in,width=2in]{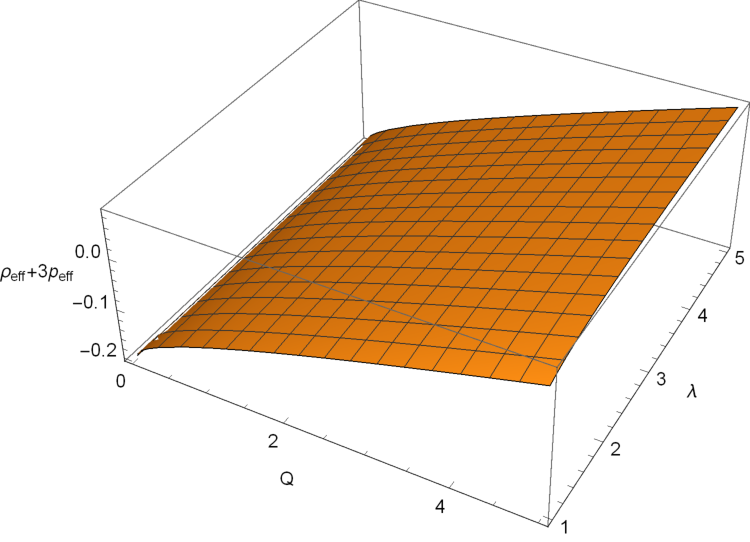}~~~~
\includegraphics[height=2in,width=2in]{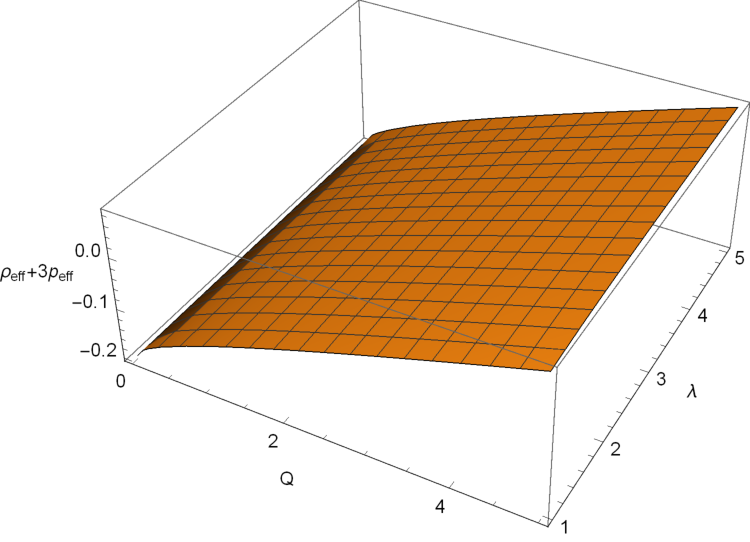}

~~~~~~~~~~~~~~~~~~(4g)~~~~~~~~~~~~~~~~~~~~~~~~~~~~~~~~~~~~~~~~~~~~~~~(4h)~~~~~~~~~~~~~~~~~~~~~~~~~~~~~~~~~~~~~~~~~~~~~~~(4i)\\

\vspace{1 mm}

\vspace{1 mm}
\end{figure}

\begin{figure}
\includegraphics[height=2in,width=2in]{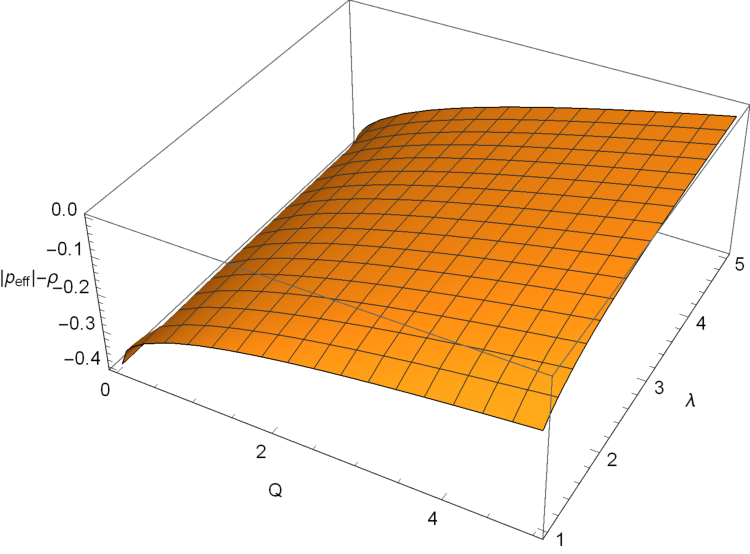}~~~~
\includegraphics[height=2in,width=2in]{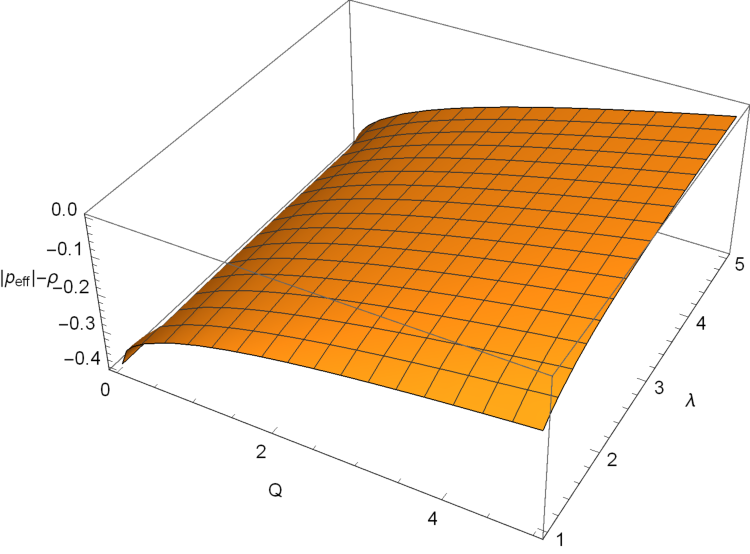}~~~~
\includegraphics[height=2in,width=2in]{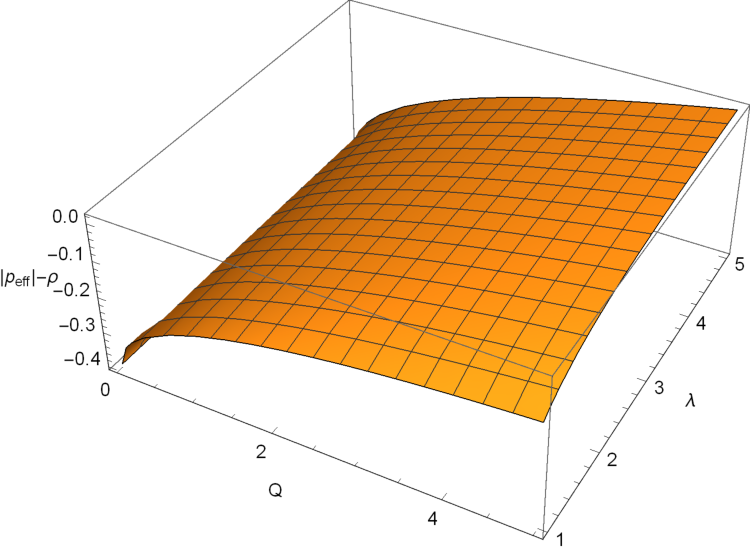}

~~~~~~~~~~~~~~~~~~(4j)~~~~~~~~~~~~~~~~~~~~~~~~~~~~~~~~~~~~~~~~~~~~~~~(4k)~~~~~~~~~~~~~~~~~~~~~~~~~~~~~~~~~~~~~~~~~~~~~~~(4l)\\

\vspace{1 mm}
\textbf{Fig.4:~(a)-(c)}: Scenario of WEC ($\rho_{eff}\geq 0$) for GO non-flat, flat and Chen-Jing models,
\textbf{Fig.4:~(d)-(f)}: Scenario of NULL ($\rho_{eff}+p_{eff}\geq 0$) for GO non-flat, flat and Chen-Jing models,
\textbf{Fig.4:~(g)-(i)}: Scenario of SEC ($\rho_{eff}+3p_{eff}\leq 0$) for GO non-flat, flat and Chen-Jing models and
\textbf{Fig.4:~(j)-(l)}: Scenario of DEC ($|p_{eff}|-\rho_{eff}\leq 0$) for GO non-flat, flat and Chen-Jing models with
the variation of $Q$ and $m$ for the different parameters like $c$, $c_{1}$, $c_{2}$, $\kappa=1$, $\alpha$, $\beta$ and $\gamma$ for \textbf{Case IV}.\\
\vspace{2 mm}
\end{figure}

\subsection{Case V:}
Finally, in this section, we explore the energy conditions for the fifth case. Using the equations (\ref{Sol13}), (\ref{Sol14}),
(\ref{Sol15}), (\ref{eff1}) and (\ref{eff2}), we get the following inequalities for WEC, NEC, DEC, SEC for the GO model:\\

\begin{itemize}
\item \textbf{WEC, NEC and DEC:}

\textbf{WEC and DEC}

\begin{equation}
\frac{c_{9}\sqrt{Q}+Q\{1-c^{2}(\alpha-(-1+q)\beta)\kappa^{2}\}-Q\Big\{1+\frac{c_{9}}{2\sqrt{Q}}-c^{2}(\alpha-(-1+q)\beta)\kappa^{2}\Big\}}{1+\frac{c_{9}}{2\sqrt{Q}}
-c^{2}(\alpha-(-1+q)\beta)\kappa^{2}}\geq 0,
\end{equation}
and\\

\textbf{WEC and NEC}

\begin{equation}
    -\frac{Q\{c_{9}(-7+q)+2\sqrt{Q}(-4+q)(1-c^{2}(\alpha+\beta-q\beta)\kappa^{2})\}}{3\{c_{9}-2\sqrt{Q}(-1+c^{2}(\alpha+\beta-q\beta)\kappa^{2})\}}\geq 0.
\end{equation}

\item \textbf{DEC:}

\begin{equation}
    -\frac{c_{9}Q}{c_{9}-2\sqrt{Q}(-1+c^{2}(\alpha+\beta-q\beta)\kappa^{2})}+\frac{1}{3}\left|Q(-4+q)\right|\leq 0.
\end{equation}

\item \textbf{SEC:}
\begin{equation}
    \frac{Q\{c_{9}(-5+q)+2\sqrt{Q}(-4+q)(1-c^{2}(\alpha+\beta-q\beta)\kappa^{2})\}}{-c_{9}+2\sqrt{Q}(-1+c^{2}(\alpha+\beta-q\beta)\kappa^{2})}\geq 0.
\end{equation}

\end{itemize}
The expressions of the energy conditions for the Chen-Jing model are given by,\\

\begin{itemize}
\item \textbf{WEC, NEC and DEC:}

\textbf{WEC and DEC}

\begin{eqnarray*}
    \Big[Q\{c_{10}+2\sqrt{6}c^{2}H(-1+q)^{2}\alpha\kappa^{2}-\sqrt{6}c^{2}H(-1+q)^{2}\alpha\kappa^{2}\ln Q\}\Big]/\Big[c_{10}+2\{-\sqrt{6}c^{2}H(-1+q)^{2}\alpha\kappa^{2}+\sqrt{Q}(1
\end{eqnarray*}
\begin{equation}
    +c^{2}((-1+q)\beta-\gamma)\kappa^{2})\}-\sqrt{6}c^{2}H(-1+q)^{2}\alpha\kappa^{2}\ln Q\Big]\geq 0,
\end{equation}
and\\

\textbf{WEC and NEC}

\begin{eqnarray*}
    -\left[Q\left\{c_{10}(-7+q)+2(-\sqrt{6}c^{2}H(-1+q)^{3}\alpha\kappa^{2}+\sqrt{Q}(-4+q)(1+c^{2}((-1+q)\beta-\gamma)\kappa^{2}))-\sqrt{6}c^{2}H(-7+q)\right.\right.
\end{eqnarray*}
\begin{eqnarray*}
    \left.\left.(-1+q)^{2}\alpha\kappa^{2}\ln
    Q\right\}\right]/\left[3\left\{c_{10}+2(-\sqrt{6}c^{2}H(-1+q)^{2}\alpha\kappa^{2}+\sqrt{Q}(1+c^{2}((-1+q)\beta-\gamma)\kappa^{2}))\right.\right.
\end{eqnarray*}

\begin{equation}
\left.\left.-\sqrt{6}c^{2}H(-1+q)^{2}\alpha\kappa^{2}\ln
Q\right\}\right]\geq 0.
\end{equation}

\item \textbf{DEC:}

\begin{eqnarray*}
\frac{1}{3}|Q(-4+q)|-\Big[Q\{c_{10}+2\sqrt{6}c^{2}H(-1+q)^{2}\alpha\kappa^{2}-\sqrt{6}c^{2}H(-1+q)^{2}\alpha\kappa^{2}\ln
Q\}\Big]/\Big[c_{10}+2\{-\sqrt{6}c^{2}H(-1+q)^{2} \alpha\kappa^{2}
\end{eqnarray*}
\begin{equation}
+\sqrt{Q}(1+c^{2}((-1+q)\beta-\gamma)\kappa^{2})\}-\sqrt{6}c^{2}H(-1+q)^{2}\alpha\kappa^{2}\ln
Q\Big]\leq 0.
\end{equation}

\item \textbf{SEC:}
\begin{eqnarray*}
-\Big[\{Q(c_{10}(-5+q)+2(-\sqrt{6}c^{2}H(-3+q)(-1+q)^{2}\alpha\kappa^{2}+\sqrt{Q}(-4+q)(1+c^{2}((-1+q)\beta-\gamma)\kappa^{2}))-\sqrt{6}c^{2}H
\end{eqnarray*}
\begin{eqnarray*}
(-5+q)(-1+q)^{2}\alpha\kappa^{2}\ln
Q)\}\Big]/\Big[c_{10}+2\{-\sqrt{6}c^{2}H(-1+q)^{2}\alpha\kappa^{2}+\sqrt{Q}(1+c^{2}((-1+q)\beta-\gamma)\kappa^{2})
\end{eqnarray*}
\begin{equation}
-\sqrt{6}c^{2}H(-1+q)^{2}\alpha\kappa^{2}\ln Q\}\Big]\geq 0.
\end{equation}

\end{itemize}

To get further understanding of these expressions, the energy conditions mentioned above are plotted in Fig.(5).\\\\

\begin{figure}
\includegraphics[height=2in,width=2in]{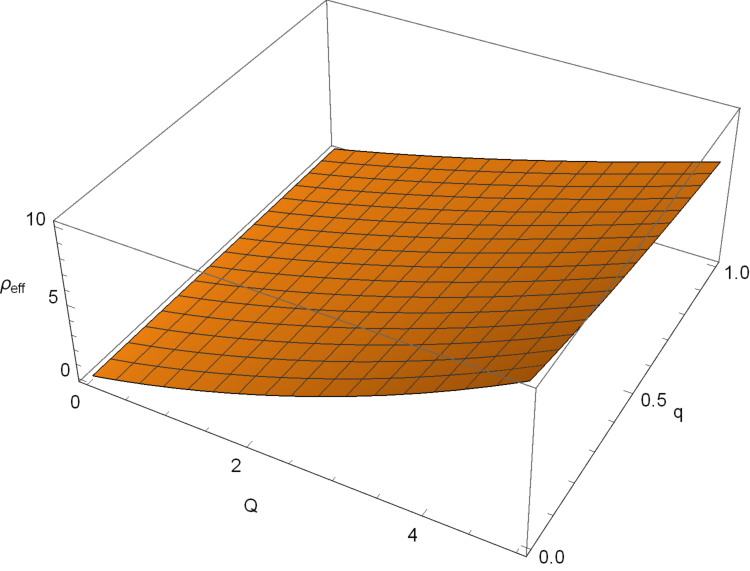}~~~~
\includegraphics[height=2in,width=2in]{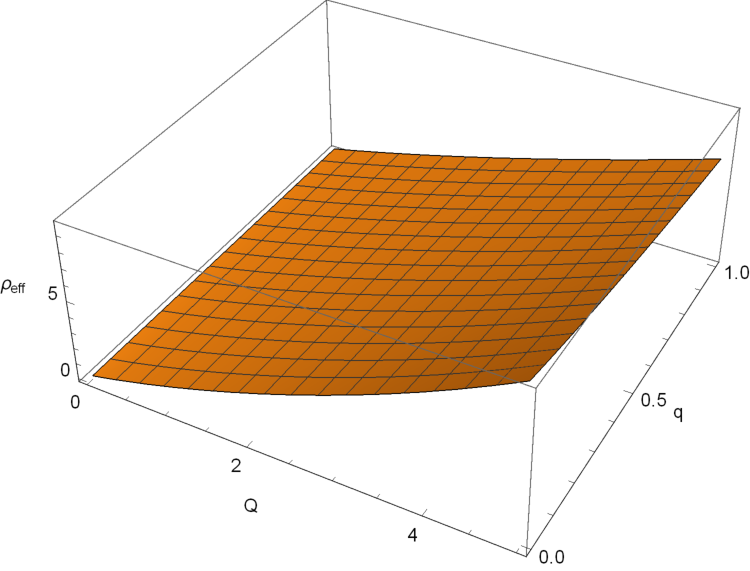}~~~~
\includegraphics[height=2in,width=2in]{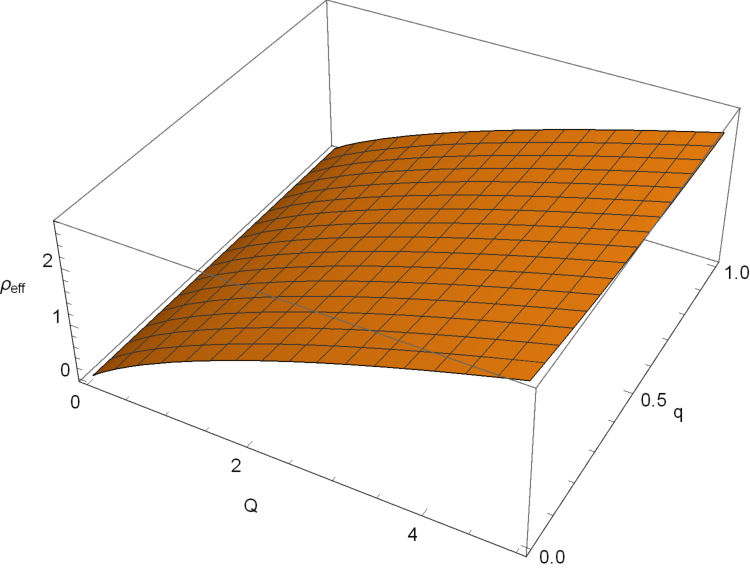}

~~~~~~~~~~~~~~~~~~(5a)~~~~~~~~~~~~~~~~~~~~~~~~~~~~~~~~~~~~~~~~~~~~~~~(5b)~~~~~~~~~~~~~~~~~~~~~~~~~~~~~~~~~~~~~~~~~~~~~~~(5c)\\

\vspace{1 mm}

\vspace{1 mm}
\end{figure}

\begin{figure}
\includegraphics[height=2in,width=2in]{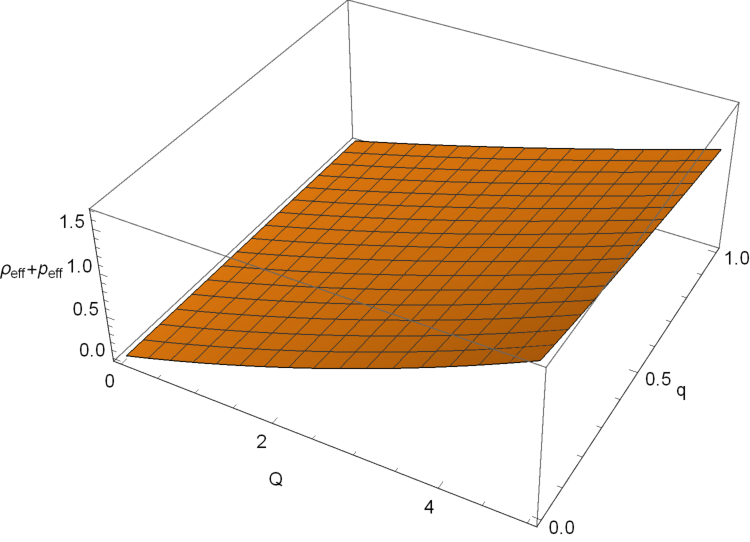}~~~~
\includegraphics[height=2in,width=2in]{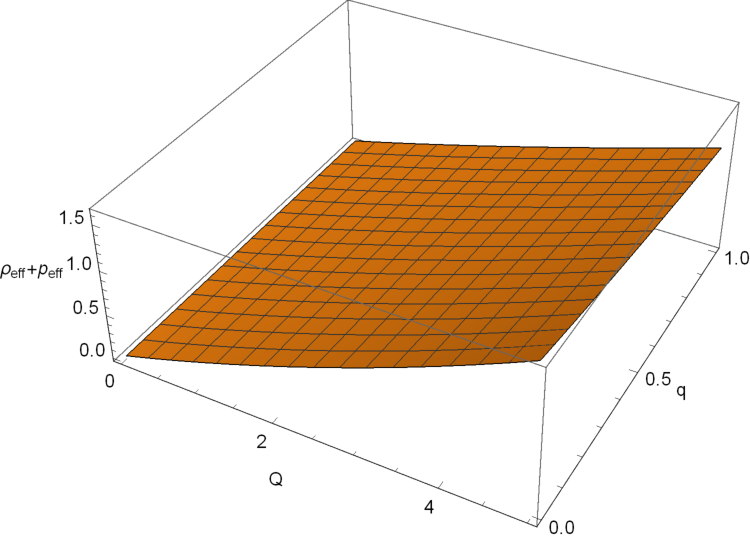}~~~~
\includegraphics[height=2in,width=2in]{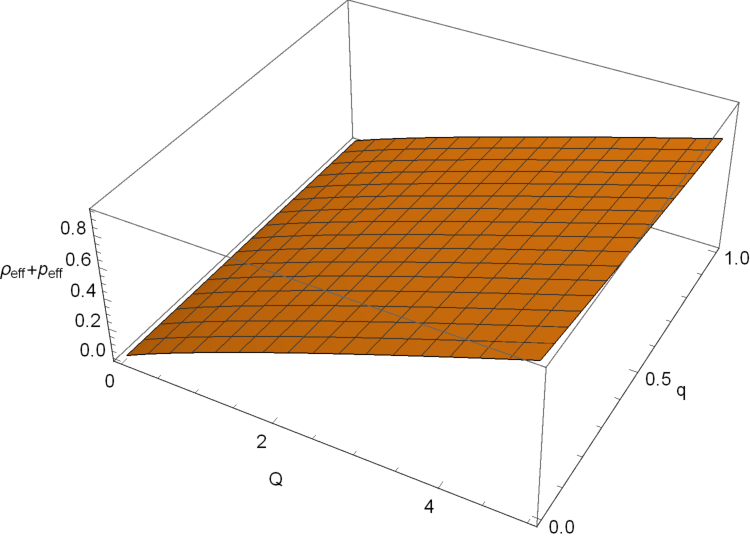}

~~~~~~~~~~~~~~~~~~(5d)~~~~~~~~~~~~~~~~~~~~~~~~~~~~~~~~~~~~~~~~~~~~~~~(5e)~~~~~~~~~~~~~~~~~~~~~~~~~~~~~~~~~~~~~~~~~~~~~~~(5f)\\

\vspace{1 mm}

\vspace{1 mm}
\end{figure}

\begin{figure}
\includegraphics[height=2in,width=2in]{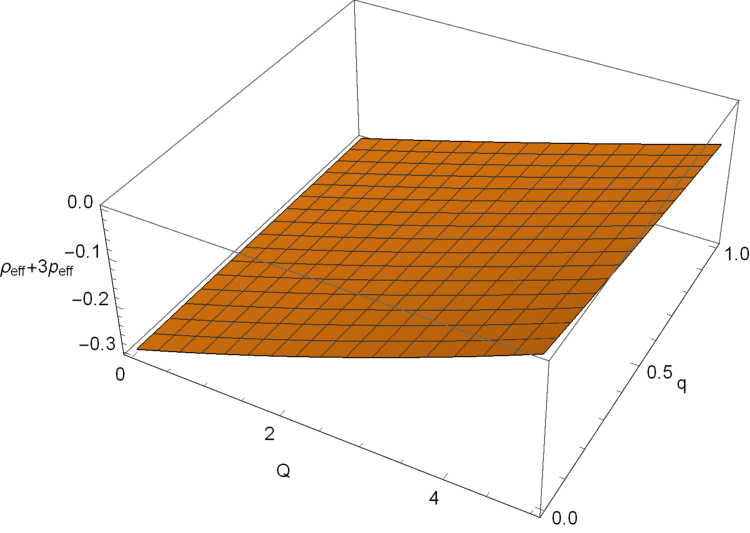}~~~~
\includegraphics[height=2in,width=2in]{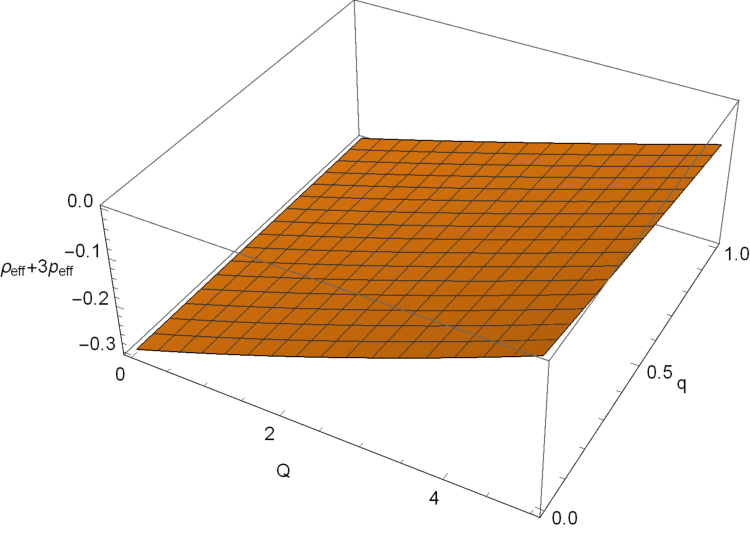}~~~~
\includegraphics[height=2in,width=2in]{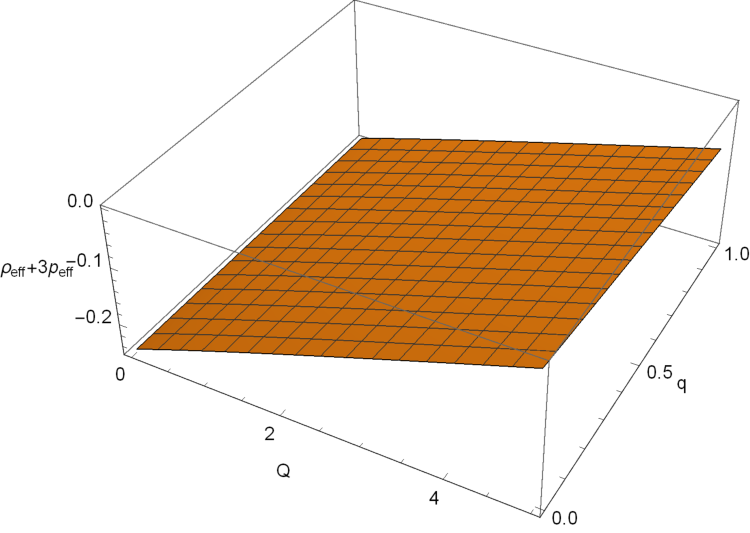}

~~~~~~~~~~~~~~~~~~(5g)~~~~~~~~~~~~~~~~~~~~~~~~~~~~~~~~~~~~~~~~~~~~~~~(5h)~~~~~~~~~~~~~~~~~~~~~~~~~~~~~~~~~~~~~~~~~~~~~~~(5i)\\

\vspace{1 mm}

\vspace{1 mm}
\end{figure}

\begin{figure}
\includegraphics[height=2in,width=2in]{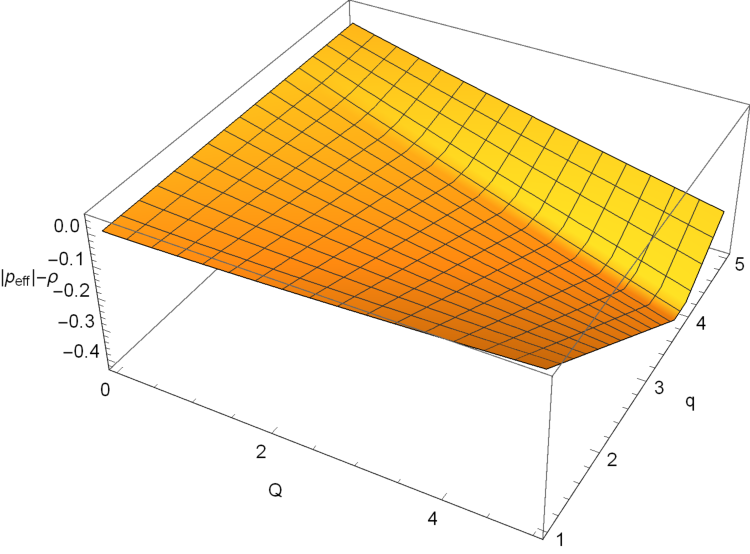}~~~~
\includegraphics[height=2in,width=2in]{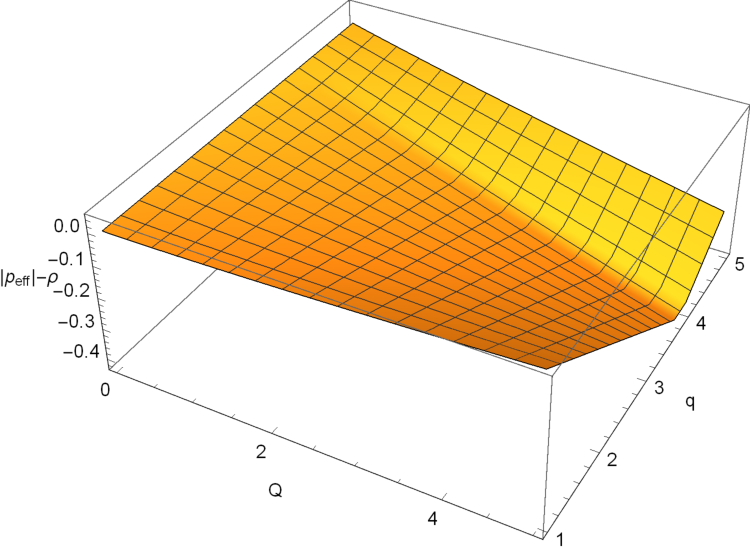}~~~~
\includegraphics[height=2in,width=2in]{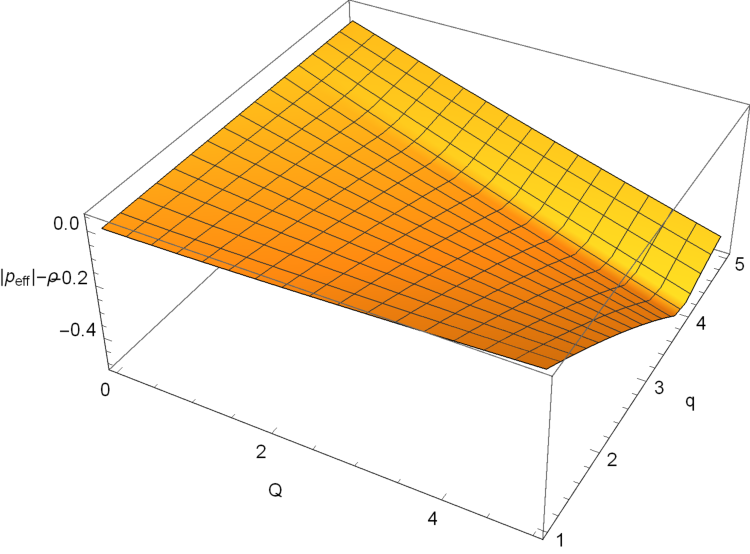}

~~~~~~~~~~~~~~~~~~(5j)~~~~~~~~~~~~~~~~~~~~~~~~~~~~~~~~~~~~~~~~~~~~~~~(5k)~~~~~~~~~~~~~~~~~~~~~~~~~~~~~~~~~~~~~~~~~~~~~~~(5l)\\

\vspace{1 mm}
\textbf{Fig.5:~(a)-(c)}: Scenario of WEC ($\rho_{eff}\geq 0$) for GO non-flat, flat and Chen-Jing models,
\textbf{Fig.5:~(d)-(f)}: Scenario of NULL ($\rho_{eff}+p_{eff}\geq 0$) for GO non-flat, flat and Chen-Jing models,
\textbf{Fig.5:~(g)-(i)}: Scenario of SEC ($\rho_{eff}+3p_{eff}\leq 0$) for GO non-flat, flat and Chen-Jing models and
\textbf{Fig.5:~(j)-(l)}: Scenario of DEC ($|p_{eff}|-\rho_{eff}\leq 0$) for GO non-flat, flat and Chen-Jing models with
the variation of $Q$ and $m$ for the different parameters like $c$, $c_{1}$, $c_{2}$, $\kappa=1$, $\alpha$, $\beta$ and $\gamma$ for \textbf{Case V}.\\
\vspace{2 mm}
\end{figure}

From all the above expressions and plots it is understandable that
we get scenarios where some energy conditions are satisfied
whereas others are not. This is expected when we work with a
modified gravity model that incorporates exotic constituents in
its framework. In particular, it is stated widely in the
literature that for dark energy or modified gravity models the
strong energy condition is generally violated. 3-D plots are
generated so that we can explore multiple parameter dependencies
for the energy conditions. From the above plots, it is quite
evident that SEC is violated in almost all the cases. We see that
the expressions for the energy conditions that have been provided
above are quite lengthy and tedious, which is expected given the
complex nature of the modified gravity. We get mixed results for
the other energy conditions depending on the case and parameter
space. These conditions aid us in investigating the potential
limits of the reconstructed $f(Q)$ models and the necessity of
developing new physics to account for phenomena observed in the
real world. These conditions act as fundamental theoretical tools
in our study that influence our knowledge of the reconstructed
$f(Q)$ models by limiting the kinds of matter and energy
distributions that are physically plausible within the context of
cosmology and the various observations related to it. In short,
these conditions help us to validate our newly reconstructed
$f(Q)$ models. We discuss the above plots in detail in the next
section and try to derive some cosmological sense out of them. It
should be noted that in the above and also in the appendix we have
presented the energy conditions in a compact manner. The common
expressions for the different energy conditions have been provided
together to avoid repetitions.

\section{Reconstruction scheme using the Generalized HDE model}
In \cite{add1} a generalized version of HDE is proposed where the
infrared cutoff is identified with the combination of FRW
parameters like the Hubble constant, particle and future horizons,
cosmological constant and universe life-time (if finite). It was
shown that depending on the specific choice of model different
interesting features occur such as the possibility to solve the
coincidence problem, crossing of phantom divide and unification of
early-time inflationary and late-time accelerating phantom
universe. In \cite{add1} the authors propose a similar generalized
HDE model where the infrared cutoff depends on the particle and
future horizons, their time derivatives and the scale factor given
below
\begin{equation}
L_{IR}=L_{IR}
(L_P,\dot{L}_{P},\ddot{L}_{P},...,L_{f},\dot{L}_{f},\ddot{L}_{f},...,a)
\end{equation}
where $L_P$ and $L_f$ are respectively the particle and the future
event horizon. An extension of this work is done in \cite{add2},
where, using this approach, it was shown that a broad class of
dark energy models can be thought of as distinct candidates for
the generalized HDE family, each with its own cut-offs. This
basically means that different dark energy models can be thought
of as special cases of this generalized HDE model. This could be
interpreted as a symmetry between several dark energy models and
the generalized HDE.

In our work, following the work done in \cite{add2}, we can find
an equivalence between the GO and the Chen-Jing HDE model with the
generalized HDE model and then reconstruct the resulting model
with the $f(Q)$ gravity to get the final results in terms of the
generalized HDE parameters. According to the holographic
principle, the inverse squared infrared cutoff $L_{IR}$ determines
the holographic energy density, which may be connected to the
causality provided by the cosmological horizon as
\begin{equation}\label{holo}
\rho_{hol}=\frac{3c^2}{\kappa^2 L_{IR}^{2}}
\end{equation}
where $\kappa^2$ is the gravitational constant and $c$ is a free
parameter. The infrared cutoff is generally taken as the particle
horizon $L_{P}$ or the future event horizon $L_{f}$ given by,
\begin{equation}
L_{P}\equiv a\int_{0}^{t}\frac{dt}{a}, ~~~~~~ L_{f}\equiv
a\int_{t}^{\infty}\frac{dt}{a}
\end{equation}
Differentiating the above expressions we get the Hubble parameter
in terms of the particle and future event horizons given by
\begin{equation}
H(L_{P},\dot{L}_{P})=\frac{\dot{L}_{P}}{L_{P}}-\frac{1}{L_{P}},
~~~~~~~~
H(L_{f},\dot{L}_{f})=\frac{\dot{L}_{f}}{L_{f}}+\frac{1}{L_{f}}
\end{equation}
Using eqns.(\ref{go2}) and (\ref{holo}) and using the above
generalized expressions for the Hubble parameter we get for the GO
model,
\begin{equation}
\frac{3c^2}{\kappa^2 L_{IR}^{2}}=3c^{2}\left[\alpha
\left(\frac{\dot{L}_{P}}{L_{P}}-\frac{1}{L_{P}}\right)^{2}+\beta\left(\frac{\ddot{L}_{P}}{L_{P}}-\frac{\dot{L}_{P}^{2}}{L_{P}^{2}}+\frac{\dot{L}_{P}}{L_{P}^{2}}\right)\right]
=3c^{2}\left[\alpha\left(\frac{\dot{L}_{f}}{L_{f}}+\frac{1}{L_{f}}\right)^{2}+\beta\left(\frac{\ddot{L}_{f}}{L_{f}}-\frac{\dot{L}_{f}^{2}}{L_{f}^{2}}-\frac{\dot{L}_{f}}{L_{f}^{2}}\right)\right]
\end{equation}
Here we have used both the definitions of the Hubble parameter
given by particle horizon and the future event horizon. Similarly
for the Chen-Jing model using eqns.(\ref{cg1}) and (\ref{holo}) we
get,

$$\frac{3c^2}{\kappa^{2}L_{IR}^{2}}=3c^{2}\left[\alpha\left\{\left(\frac{\dddot{L}_{P}}{L_{P}}-3\frac{\dot{L}_{P}\ddot{L}_{P}}{L_{P}^{2}}+\frac{\ddot{L}_{P}}{L_{P}^{2}}-2\frac{\dot{L}_{P}^{2}}{L_{P}^{3}}
\left(1-\dot{L}_{P}\right)\right)\left(\frac{\dot{L}_{P}}{L_{P}}-\frac{1}{L_{P}}\right)^{-1}\right\}+\beta\left(\frac{\ddot{L}_{P}}{L_{P}}-\frac{\dot{L}_{P}^{2}}{L_{P}^{2}}
+\frac{\dot{L}_{P}}{L_{P}^{2}}\right)\right.$$

$$\left.+\gamma\left(\frac{\dot{L}_{P}}{L_{P}}-\frac{1}{L_{P}}\right)^{2}\right]=3c^{2}\left[\alpha\left\{\left(\frac{\dddot{L}_{f}}{L_{f}}-3\frac{\dot{L}_{f}\ddot{L}_{f}}{L_{f}^{2}}-\frac{\ddot{L}_{f}}{L_{f}^{2}}+2\frac{\dot{L}_{f}^{2}}{L_{f}^{3}}
\left(1+\dot{L}_{f}\right)\right)\left(\frac{\dot{L}_{f}}{L_{f}}+\frac{1}{L_{f}}\right)^{-1}\right\}\right.$$

\begin{equation}
\left.+\beta\left(\frac{\ddot{L}_{f}}{L_{f}}-\frac{\dot{L}_{f}^{2}}{L_{f}^{2}}
-\frac{\dot{L}_{f}}{L_{f}^{2}}\right)+\gamma\left(\frac{\dot{L}_{f}}{L_{f}}+\frac{1}{L_{f}}\right)^{2}\right]
\end{equation}

The above expressions give the equivalent forms of the GO and
Chen-Jing models in terms of the generalized HDE parameters. Now
using the $f(Q)$ dark energy density from eqn.(\ref{rhode1}) with
the above newly formed expressions we can easily set up a
reconstruction scheme to get the reconstructed $f(Q)$ models.

%%%%%%%%%%%%%%%%%%%%%%%%%%%%%%%%%%%%%
\section{Conclusion and Discussion}
%%%%%%%%%%%%%%%%%%%%%%%%%%%%%%%%%%%%%
In this paper, we have explored a reconstruction scheme of $f(Q)$
gravity using holographic dark energy. Two different holographic
dark energy models have been used namely, Granda-Oliveros dark
energy and Chen-Jing dark energy, and a direct correspondence is
set up with $f(Q)$ gravity. The reconstructed $f(Q)$ models are
formed using this correspondence for both cases. For the
Granda-Oliveros case, we have used observationally motivated
values of the model parameters. As stated earlier, the goal of
this cosmological reconstruction approach is to use modified
gravity theories to precisely recover the $\Lambda$CDM features
and determine the universe's expansion history.

In order to check the viability of the the reconstructed $f(Q)$
models we have investigated the energy conditions for the models.
Four different energy conditions, namely the weak energy
condition, null energy condition, dominant energy condition, and
strong energy condition are explored for the reconstructed
modified $f(Q)$ gravity models. The results have been plotted in
Figs.(1), (2), (3), (4), and (5) for the different cases of scale
factors. In Fig.(1) we have plotted the energy conditions for the
reconstructed model obtained for the scale factor in case I. In
Figs. I(a), I(b), and I(c) we have plotted the WEC for the GO
non-flat, flat, and the Chen-Jing model respectively. The plot is
generated against the non-metricity $Q$ and the model parameter
$m$. The other parameters are considered as unity. It can be
clearly seen in the plots that the condition is satisfied in the
given range of the parameters. In Figs. I(d), I(e), and I(f)
similar plots are generated for NEC. From the plots, we see that
for some range of the parameter, there may be a violation of the
NEC, which is not unexpected for any exotic component generated
from an alternative theory of gravity. In Figs.I(g), I(h), and
I(i) we have the corresponding plots for SEC. From the plots, we
see that there is a clear violation of the strong energy condition
in some ranges of the free parameters for all the cases. This is
attributed to the anti-gravitating stress generated by the exotic
component generated from the reconstructed modified gravity
models. Finally in Figs.I(j), I(k), and I(l) the results for DEC
are plotted. In the figures, we see that DEC is satisfied for all
the cases. In Figs.2(a)-(l) similar plots are generated for the
reconstructed models for the scale factor discussed in case II.
From the plots, we see that except SEC (which is violated) the
other energy conditions are satisfied for the given range of
parameters. In Figs.3(a)-(l) similar plots are generated for the
reconstructed model for the scale factor discussed in case III.
Here also see results, similar to those of the previous case. In
Figs.4(a)-(l) similar plots are generated for the reconstructed
models for the scale factor discussed in case IV. Here also we see
that there is a violation of SEC while the other energy conditions
are satisfied. In Figs.5(a)-(l) corresponding plots are developed
for the reconstructed models for the scale factor discussed in
case V. Except for little variation at some places the plots show
a similar tendency as compared to the the previous cases.

So from the above discussion, it is understandable that the
reconstructed $f(Q)$ models are good candidates for modified
gravity theory. The violation of the SEC for all the cases
considered above indicate that the reconstructed models possess an
exotic tendency when compared to a dark energy model. This is a
necessary condition for the models to be able to drive the late
cosmic acceleration. Moreover, since the $f(Q)$ models are
reconstructed from holographic dark energy models (having origin
in black hole thermodynamics), these models also have the
potential to successfully model the early inflationary phase.
These reconstructed models are backed by the observational data
and hence are constrained. So the models of $f(Q)$ gravity
produced from this study are potentially viable models to describe
both the early time inflation and the late time acceleration.
Further studies like dynamical system analysis, and perturbation
study need to be conducted to get deeper insights into the
reconstructed models. But as of now, we have some potentially good
candidates to
model the entire evolution of the universe.\\

\section*{Appendix}
Since the expressions for the energy conditions of Case-III and
Case-IV are very lengthy we report them here instead of the main
body of the paper.

\textbf{Energy Conditions for Case III and Case IV}

\subsection*{Case III:}
Using the equations (\ref{Sol7}), (\ref{Sol8}), (\ref{Sol9}), (\ref{eff1}) and (\ref{eff2}), we get the following inequalities for WEC, NEC, DEC, SEC for the GO model:
\begin{itemize}

\item \textbf{WEC, NEC and DEC:}

\item \textbf{WEC and DEC}

\begin{eqnarray*}
    [Q\{6^{1-\frac{3\theta}{2}}(6^{1+\frac{\theta^{2}}{2}}c_{5}\sqrt{Q}+Q^{\frac{1}{2}-\frac{3\theta}{2}}(B\theta)^{\frac{1}{-1+\theta}}(-6^{\frac{3\theta}{2}}c^{2}Q^{\frac{1+\theta^{2}}{2}}\beta(-3+\theta)(-1+\theta)^{2}\theta+2^{\frac{3+\theta^{2}}{2}}\times3^{\frac{1+\theta^{2}}{2}}BQ^{\frac{3\theta}{2}}\beta nc^{2}(B\theta)^{\frac{1}{1-\theta}})
\end{eqnarray*}
\begin{eqnarray*}
    \kappa^{2})-6^{\frac{3-3\theta+\theta^{2}}{2}}B\sqrt{Q}\beta nc^{2}\kappa^{2}\ln Q\}]/[6^{1-\frac{3\theta}{2}}\{6^{1+\frac{\theta^{2}}{2}}c_{5}\sqrt{Q}-2^{\frac{3+\theta^{2}}{2}}.3^{\frac{1+\theta^{2}}{2}}B\sqrt{Q}\beta nc^{2}\kappa^{2}+6^{3\theta/2}c^{2}Q^{\frac{1}{2}(-2+\theta)(-1+\theta)}
\end{eqnarray*}
\begin{equation}
    \beta(-2+\theta)(-1+\theta)^{3}(B\theta)^{\frac{1}{\theta-1}}\kappa^{2}-2^{2+\theta^{2}/2}\times3^{1+\theta^{2}/2}Q(-1+c^{2}(\alpha-\beta)\kappa^{2})\}-6^{\frac{3-3\theta+\theta^{2}}{2}}B\sqrt{Q}\beta nc^{2}\kappa^{2}\ln Q]\geq 0,
\end{equation}
and\\
\item \textbf{WEC and NEC}

\begin{eqnarray*}
    \Big[2Q\Big\{6c_{5}+12\sqrt{Q}-12c^{2}\sqrt{Q}(\alpha-\beta)\kappa^{2}-2\sqrt{6}B\beta nc^{2}\kappa^{2}+6^{\frac{-(-3+\theta)\theta}{2}}c^{2}Q^{\frac{1-3\theta+\theta^{2}}{2}}\beta(-2+\theta)(-1+\theta)^{3}(B\theta)^{\frac{1}{-1+\theta}}\kappa^{2}
\end{eqnarray*}
\begin{eqnarray*}
    -\sqrt{6}B\beta nc^{2}\kappa^{2}\ln Q\Big\}-6^{\frac{\theta}{2(-1+\theta)}}B\Big(\frac{\sqrt{Q}}{B\theta}\Big)^{\frac{-2+\theta}{-1+\theta}}\theta\Big\{1-\Big(1+6^{\frac{\theta}{2-2\theta}}B\Big(\frac{\sqrt{Q}}{B\theta}\Big)^{\frac{\theta}{-1+\theta}}\Big)\theta\Big\}\{6c_{5}+12\sqrt{Q}-12c^{2}\sqrt{Q}(\alpha-\beta)\kappa^{2}
\end{eqnarray*}
\begin{eqnarray*}
    -2\sqrt{6}B\beta nc^{2}\kappa^{2}+6^{\frac{-(-3+\theta)\theta}{2}}c^{2}Q^{\frac{1-3\theta+\theta^{2}}{2}}\beta(-2+\theta)(-1+\theta)^{3}(B\theta)^{\frac{1}{-1+\theta}}\kappa^{2}-\sqrt{6}B\beta nc^{2}\kappa^{2}\ln Q\}+2^{\frac{-(-2+\theta)(-1+\theta)}{2}}
\end{eqnarray*}
\begin{eqnarray*}
    \times3^{\frac{-(-3+\theta)\theta}{2}}\Big\{6^{1-3\theta/2}Q^{1-3\theta/2}\Big(6^{1+\theta^{2}/2}c_{5}Q^{3\theta/2}+(B\theta)^{\frac{1}{-1+\theta}}\Big(-6^{3\theta/2}c^{2}Q^{\frac{1+\theta^{2}}{2}}\beta(-3+\theta)(-1+\theta)^{2}\theta+2^{\frac{3+\theta^{2}}{2}}
\end{eqnarray*}
\begin{eqnarray*}
    \times3^{\frac{1+\theta^{2}}{2}}BQ^{3\theta/2}\beta nc^{2}(B\theta)^{\frac{1}{1-\theta}}\Big)\kappa^{2}\Big)-6^{\frac{3-3\theta+\theta^{2}}{2}}BQ\beta nc^{2}\kappa^{2}\ln Q\Big\}\Big]/\Big[3\sqrt{Q}\Big\{12+\frac{6c_{5}}{\sqrt{Q}}-12c^{2}(\alpha-\beta)\kappa^{2}-\frac{2\sqrt{6}B\beta nc^{2}\kappa^{2}}{\sqrt{Q}}
\end{eqnarray*}
\begin{equation}
    +6^{\frac{-(-3+\theta)\theta}{2}}c^{2}Q^{\frac{(-3+\theta)\theta}{2}}\beta(-2+\theta)(-1+\theta)^{3}(B\theta)^{\frac{1}{-1+\theta}}\kappa^{2}-\frac{\sqrt{6}B\beta nc^{2}\kappa^{2}\ln Q}{\sqrt{Q}}\Big\}\Big]\geq 0.
\end{equation}

\item \textbf{DEC:}

\begin{eqnarray*}
    \frac{1}{3}\exp^{-2Re\Big[\frac{\ln \Big[\frac{\sqrt{Q}}{B\theta}\Big]}{-1+\theta}\Big]}\Big|2Q\Big(\frac{\sqrt{Q}}{B\theta}\Big)^{\frac{2}{-1+\theta}}+B\Big(\frac{\sqrt{Q}}{B\theta}\Big)^{\frac{\theta}{-1+\theta}}\theta\Big\{-6^{\frac{\theta}{2(-1+\theta}}+\Big(6^{\frac{\theta}{2(-1+\theta}}+B\Big(\frac{\sqrt{Q}}{B\theta}\Big)^{\frac{\theta}{-1+\theta}}\Big)\theta\Big\}\Big|
\end{eqnarray*}
\begin{eqnarray*}
    -\Big[Q\Big\{6^{1-3\theta/2}\Big(6^{1+\theta^{2}/2}c_{5}\sqrt{Q}+Q^{\frac{1}{2}-\frac{3\theta}{2}}(B\theta)^{\frac{1}{-1+\theta}}\Big(-6^{3\theta/2}c^{2}Q^{\frac{1+\theta^{2}}{2}}\beta(-3+\theta)(-1+\theta)^{2}\theta
\end{eqnarray*}
\begin{eqnarray*}
    +2^{\frac{3+\theta^{2}}{2}}\times3^{\frac{1+\theta^{2}}{2}}BQ^{3\theta/2}\beta nc^{2}(B\theta)^{\frac{1}{1-\theta}}\Big)\kappa^{2}\Big)-6^{\frac{3-3\theta+\theta^{2}}{2}}B\sqrt{Q}\beta nc^{2}\kappa^{2}\ln Q\Big\}\Big]/\Big[6^{1-3\theta/2}\Big\{6^{1+\theta^{2}/2}c_{5}\sqrt{Q}
\end{eqnarray*}
\begin{eqnarray*}
    -2^{\frac{3+\theta^{2}}{2}}\times3^{\frac{1+\theta^{2}}{2}}B\sqrt{Q}\beta nc^{2}\kappa^{2}+6^{3\theta/2}c^{2}Q^{\frac{(-2+\theta)(-1+\theta)}{2}}\beta(-2+\theta)(-1+\theta)^{3}(B\theta)^{\frac{1}{-1+\theta}}\kappa^{2}
\end{eqnarray*}
\begin{equation}
    -2^{2+\theta^{2}/2}\times3^{1+\theta^{2}/2}Q(-1+c^{2}(\alpha-\beta)\kappa^{2})\Big\}-6^{\frac{3-3\theta+\theta^{2}}{2}}B\sqrt{Q}\beta nc^{2}\kappa^{2}\ln Q\Big]\leq 0.
\end{equation}

\item \textbf{SEC:}
\begin{eqnarray*}
    \Big[2Q\Big\{6c_{5}+12\sqrt{Q}-12c^{2}\sqrt{Q}(\alpha-\beta)\kappa^{2}-2\sqrt{6}B\beta nc^{2}\kappa^{2}+6^{\frac{-(-3+\theta)\theta}{2}}c^{2}Q^{\frac{1-3\theta+\theta^{2}}{2}}\beta(-2+\theta)(-1+\theta)^{3}(B\theta)^{\frac{1}{-1+\theta}}\kappa^{2}
\end{eqnarray*}
\begin{eqnarray*}
    -\sqrt{6}B\beta nc^{2}\kappa^{2}\ln Q\Big\}-6^{\frac{\theta}{2(-1+\theta)}}B\Big(\frac{\sqrt{Q}}{B\theta}\Big)^{\frac{-2+\theta}{-1+\theta}}\theta\Big\{1-\Big(1+6^{\frac{\theta}{2-2\theta}}B\Big(\frac{\sqrt{Q}}{B\theta}\Big)^{\frac{\theta}{-1+\theta}}\Big)\theta\Big\}\Big\{6c_{5}+12\sqrt{Q}-12c^{2}\sqrt{Q}(\alpha-\beta)\kappa^{2}
\end{eqnarray*}
\begin{eqnarray*}
    -2\sqrt{6}B\beta nc^{2}\kappa^{2}+6^{\frac{-(-3+\theta)\theta}{2}}c^{2}Q^{\frac{1-3\theta+\theta^{2}}{2}}\beta(-2+\theta)(-1+\theta)^{3}(B\theta)^{\frac{1}{-1+\theta}}\kappa^{2}-\sqrt{6}B\beta nc^{2}\kappa^{2}\ln Q\Big\}+6^{\frac{-(-2+\theta)(-1+\theta)}{2}}
\end{eqnarray*}
\begin{eqnarray*}
    \Big\{6^{1-3\theta/2}Q^{1-3\theta/2}\Big(6^{1+\theta^{2}/2}c_{5}Q^{3\theta/2}+(B\theta)^{\frac{1}{-1+\theta}}\Big(-6^{3\theta/2}c^{2}Q^{\frac{1+\theta^{2}}{2}}\beta(-3+\theta)(-1+\theta)^{2}\theta+2^{\frac{3+\theta^{2}}{2}}\times3^{\frac{1+\theta^{2}}{2}}BQ^{3\theta/2}
\end{eqnarray*}
\begin{eqnarray*}
    \beta nc^{2}(B\theta)^{\frac{1}{1-\theta}}\Big)\kappa^{2}\Big)-6^{\frac{3-3\theta+\theta^{2}}{2}}BQ\beta nc^{2}\kappa^{2}\ln Q\Big\}\Big]/\Big\{6c_{5}+12\sqrt{Q}-12c^{2}\sqrt{Q}(\alpha-\beta)\kappa^{2}-2\sqrt{6}B\beta nc^{2}\kappa^{2}+6^{\frac{-(-3+\theta)\theta}{2}}c^{2}
\end{eqnarray*}
\begin{equation}
    Q^{\frac{1-3\theta+\theta^{2}}{2}}\beta(-2+\theta)(-1+\theta)^{3}(B\theta)^{\frac{1}{-1+\theta}}\kappa^{2}-\sqrt{6}B\beta nc^{2}\kappa^{2}\ln Q\Big\}\geq 0.
\end{equation}

\end{itemize}
The corresponding conditions for the  Chen-Jing model are given as,\\

\begin{itemize}
    \item \textbf{WEC, NEC and DEC:}

    \textbf{WEC and DEC}

    \begin{eqnarray*}
        \Big[-c_{6}\sqrt{Q}(-3+\theta)-2^{3+\frac{1}{1-\theta}}.3^{\frac{-2+\theta}{-1+\theta}}c^{2}Q^{\frac{1}{-1+\theta}}\alpha\kappa^{2}-2c^{2}Q(2+\gamma)(-3+\theta)\kappa^{2}+2^{3+\frac{1}{1-\theta}}\times3^{\frac{-2+\theta}{-1+\theta}}c^{2}Q^{\frac{1}{-1+\theta}}\alpha(-1+\theta)\kappa^{2}
    \end{eqnarray*}
    \begin{eqnarray*}
        -2^{\frac{2+3\theta-\theta^{2}}{2}}\times3^{\frac{-(-3+\theta)\theta}{2}}c^{2}Q^{\frac{(-2+\theta)(-1+\theta)}{2}}\beta(-3+\theta)(-1+\theta)^{2}(B\theta)^{\frac{1}{-1+\theta}}\kappa^{2}+6^{\frac{-(-3+\theta)\theta}{2}}c^{2}Q^{\frac{(-2+\theta)(-1+\theta)}{2}}\beta(-3+\theta)
    \end{eqnarray*}
    \begin{eqnarray*}
        (-2+\theta)(-1+\theta)^{3}(B\theta)^{\frac{1}{-1+\theta}}\kappa^{2}+9\sqrt{6}c^{2}\sqrt{\frac{1}{Q}}\alpha(-3+\theta)(-2+\theta)(-1+\theta)(B\theta)^{\frac{2}{-1+\theta}}\kappa^{2}+2Q(3-\theta)\{1-c^{2}(2+\gamma)\kappa^{2}\}\Big]
    \end{eqnarray*}
    \begin{eqnarray*}
        /\Big[(3-\theta)\Big\{2+\frac{c_{6}}{\sqrt{Q}}-2c^{2}(2+\gamma)\kappa^{2}-\frac{2^{3+\frac{1}{1-\theta}}\times3^{\frac{-2+\theta}{-1+\theta}}c^{2}Q^{-1+\frac{1}{-1+\theta}}\alpha\kappa^{2}}{-3+\theta}+6^{\frac{-(-3+\theta)\theta}{2}}c^{2}Q^{\frac{(-3+\theta)\theta}{2}}\beta(-2+\theta)(-1+\theta)^{3}
    \end{eqnarray*}
    \begin{equation}
        (B\theta)^{\frac{1}{-1+\theta}}\kappa^{2}+3\sqrt{6}c^{2}\Big(\frac{1}{Q}\Big)^{3/2}\alpha(-2+\theta)(-1+\theta)(B\theta)^{\frac{2}{-1+\theta}}\kappa^{2}\Big\}\Big]\geq 0.
    \end{equation}
and\\
\textbf{WEC and NEC}
\begin{eqnarray*}
    \Big[(1-\theta)\Big\{2Q\Big(-c_{6}\sqrt{Q}(-3+\theta)-2Q(-3+\theta)+2^{3+\frac{1}{1-\theta}}\times3^{\frac{-2+\theta}{-1+\theta}}c^{2}Q^{\frac{1}{-1+\theta}}\alpha\kappa^{2}+2c^{2}Q(2+\gamma)(-3+\theta)\kappa^{2}
\end{eqnarray*}
\begin{eqnarray*}
    -6^{\frac{-(-3+\theta)\theta}{2}}c^{2}Q^{\frac{(-2+\theta)(-1+\theta)}{2}}\beta(-3+\theta)(-2+\theta)(-1+\theta)^{3}(B\theta)^{\frac{1}{-1+\theta}}\kappa^{2}-3\sqrt{6}c^{2}\sqrt{\frac{1}{Q}}\alpha(-3+\theta)(-2+\theta)(-1+\theta)
\end{eqnarray*}
\begin{eqnarray*}
    (B\theta)^{\frac{2}{-1+\theta}}\kappa^{2}\Big\}-6^{\frac{\theta}{2(-1+\theta)}}B\Big(\frac{\sqrt{Q}}{B\theta}\Big)^{\frac{-2+\theta}{-1+\theta}}\theta\Big\{1-\Big(1+6^{\frac{\theta}{2-2\theta}}B\Big(\frac{\sqrt{Q}}{B\theta}\Big)^{\frac{\theta}{-1+\theta}}\Big)\theta\Big\}\Big\{-c_{6}\sqrt{Q}(-3+\theta)-2Q(-3+\theta)
\end{eqnarray*}
\begin{eqnarray*}
    +2^{3+\frac{1}{1-\theta}}\times3^{\frac{-2+\theta}{-1+\theta}}c^{2}Q^{\frac{1}{-1+\theta}}\alpha\kappa^{2}+2c^{2}Q(2+\gamma)(-3+\theta)\kappa^{2}-6^{\frac{-(-3+\theta)\theta}{2}}c^{2}Q^{\frac{(-2+\theta)(-1+\theta)}{2}}\beta(-3+\theta)(-2+\theta)(-1+\theta)^{3}
\end{eqnarray*}
\begin{eqnarray*}
    (B\theta)^{\frac{1}{-1+\theta}}\kappa^{2}-3\sqrt{6}c^{2}\sqrt{\frac{1}{Q}}\alpha(-3+\theta)(-2+\theta)(-1+\theta)(B\theta)^{\frac{2}{-1+\theta}}\kappa^{2}\Big\}+3Q\Big\{-c_{6}\sqrt{Q}(-3+\theta)+2Q(-3+\theta)
\end{eqnarray*}
\begin{eqnarray*}
    -2^{3+\frac{1}{1-\theta}}\times3^{\frac{-2+\theta}{-1+\theta}}c^{2}Q^{\frac{1}{-1+\theta}}\alpha\kappa^{2}-2c^{2}Q(2+\gamma)(-3+\theta)\kappa^{2}+2^{3+\frac{1}{1-\theta}}\times3^{\frac{-2+\theta}{-1+\theta}}c^{2}Q^{\frac{1}{-1+\theta}}\alpha(-1+\theta)\kappa^{2}
\end{eqnarray*}
\begin{eqnarray*}
    -2^{\frac{2+3\theta-\theta^{2}}{2}}\times3^{\frac{-(-3+\theta)\theta}{2}}c^{2}Q^{\frac{(-2+\theta)(-1+\theta)}{2}}\beta(-3+\theta)(-1+\theta)^{2}(B\theta)^{\frac{1}{-1+\theta}}\kappa^{2}+6^{\frac{-(-3+\theta)\theta}{2}}c^{2}Q^{\frac{(-2+\theta)(-1+\theta)}{2}}
\end{eqnarray*}
\begin{eqnarray*}
    \beta(-3+\theta)(-2+\theta)(-1+\theta)^{3}(B\theta)^{\frac{1}{-1+\theta}}\kappa^{2}+9\sqrt{6}c^{2}\sqrt{\frac{1}{Q}}\alpha(-3+\theta)(-2+\theta)(-1+\theta)(B\theta)^{\frac{2}{-1+\theta}}\kappa^{2}
\end{eqnarray*}
\begin{eqnarray*}
    +2Q(3-\theta)(1-c^{2}(2+\gamma)\kappa^{2})\Big\}\Big]/\Big[3Q(-3+\theta)(-1+\theta)\Big\{2+\frac{c_{6}}{\sqrt{Q}}-2c^{2}(2+\gamma)\kappa^{2}-\frac{2^{3+\frac{1}{1-\theta}}\times3^{\frac{-2+\theta}{-1+\theta}}c^{2}Q^{-1+\frac{1}{-1+\theta}}\alpha\kappa^{2}}{-3+\theta}
\end{eqnarray*}
\begin{equation}
    +6^{\frac{-(-3+\theta)\theta}{2}}c^{2}Q^{\frac{(-3+\theta)\theta}{2}}\beta(-2+\theta)(-1+\theta)^{3}(B\theta)^{\frac{1}{-1+\theta}}\kappa^{2}+3\sqrt{6}c^{2}\Big(\frac{1}{Q}\Big)^{3/2}\alpha(-2+\theta)(-1+\theta)(B\theta)^{\frac{2}{-1+\theta}}\kappa^{2}\Big\}\Big]\geq 0.
\end{equation}

\item \textbf{DEC:}

\begin{eqnarray*}
    \Big[\Big\{-c_{6}\sqrt{Q}(-3+\theta)+2Q(-3+\theta)-2^{3+\frac{1}{1-\theta}}\times3^{\frac{-2+\theta}{-1+\theta}}c^{2}Q^{\frac{1}{-1+\theta}}\alpha\kappa^{2}-2c^{2}Q(2+\gamma)(-3+\theta)\kappa^{2}
\end{eqnarray*}
\begin{eqnarray*}
    +2^{3+\frac{1}{1-\theta}}\times3^{\frac{-2+\theta}{-1+\theta}}c^{2}Q^{\frac{1}{-1+\theta}}\alpha(-1+\theta)\kappa^{2}-2^{\frac{2+3\theta-\theta^{2}}{2}}\times3^{\frac{-(-3+\theta)\theta}{2}}c^{2}Q^{\frac{(-2+\theta)(-1+\theta)}{2}}\beta(-3+\theta)(-1+\theta)^{2}(B\theta)^{\frac{1}{-1+\theta}}\kappa^{2}
\end{eqnarray*}
\begin{eqnarray*}
    +6^{\frac{-(-3+\theta)\theta}{2}}c^{2}Q^{\frac{(-2+\theta)(-1+\theta)}{2}}\beta(-3+\theta)(-2+\theta)(-1+\theta)^{3}(B\theta)^{\frac{1}{-1+\theta}}\kappa^{2}+9\sqrt{6}c^{2}\sqrt{\frac{1}{Q}}\alpha(-3+\theta)(-2+\theta)(-1+\theta)
\end{eqnarray*}
\begin{eqnarray*}
    (B\theta)^{\frac{2}{-1+\theta}}\kappa^{2}+2Q(3-\theta)\{1-c^{2}(2+\gamma)\kappa^{2}\}\Big]/\Big[(3-\theta)\Big\{2+\frac{c_{6}}{\sqrt{Q}}-2c^{2}(2+\gamma)\kappa^{2}-\frac{2^{3+\frac{1}{1-\theta}}\times3^{\frac{-2+\theta}{-1+\theta}}c^{2}Q^{-1+\frac{1}{-1+\theta}}\alpha\kappa^{2}}{-3+\theta}
\end{eqnarray*}
\begin{eqnarray*}
    +6^{\frac{-(-3+\theta)\theta}{2}}c^{2}Q^{\frac{(-3+\theta)\theta}{2}}\beta(-2+\theta)(-1+\theta)^{3}(B\theta)^{\frac{1}{-1+\theta}}\kappa^{2}+3\sqrt{6}c^{2}\Big(\frac{1}{Q}\Big)^{3/2}\alpha(-2+\theta)(-1+\theta)(B\theta)^{\frac{1}{-1+\theta}}\kappa^{2}\Big\}\Big\}\Big]
\end{eqnarray*}
\begin{equation}
    +\frac{1}{3}\exp^{-2Re\Big[\frac{\ln\Big[\frac{\sqrt{Q}}{B\theta}\Big]}{-1+\theta}\Big]}\Big|2Q\Big(\frac{\sqrt{Q}}{B\theta}\Big)^{\frac{2}{-1+\theta}}+B\Big(\frac{\sqrt{Q}}{B\theta}\Big)^{\frac{\theta}{-1+\theta}}\theta\Big\{-6^{\frac{\theta}{2(-1+\theta}}+\Big(6^{\frac{\theta}{2(-1+\theta}}+B\Big(\frac{\sqrt{Q}}{B\theta}\Big)^{\frac{\theta}{-1+\theta}}\Big)\theta\Big\}\Big|\geq 0.
\end{equation}

\item \textbf{SEC:}
\begin{eqnarray*}
    \left[2Q(1-\theta)\left(-c_{6}\sqrt{Q}(-3+\theta)-2Q(-3+\theta)+2^{3+\frac{1}{1-\theta}}\times3^{\frac{-2+\theta}{-1+\theta}}c^{2}Q^{\frac{1}{-1+\theta}}\alpha\kappa^{2}+2c^{2}
    Q(2+\gamma)(-3+\theta)\kappa^{2}\right.\right.
\end{eqnarray*}
\begin{eqnarray*}
   \left.\left.-6^{\frac{-(-3+\theta)\theta}{2}}c^{2}
   Q^{\frac{(-2+\theta)(-1+\theta)}{2}}\beta(-3+\theta)(-2+\theta)(-1+\theta)^{3}(B\theta)^{\frac{1}{-1+\theta}}\kappa^{2}-3\sqrt{6}c^{2}\sqrt{\frac{1}{Q}}
   \alpha(-3+\theta)(-2+\theta)(-1+\theta)\right.\right.
\end{eqnarray*}
\begin{eqnarray*}
    \left.\left.(B\theta)^{\frac{2}{-1+\theta}}\kappa^{2}\right)-6^{\frac{\theta}{2(-1+\theta)}}B(1-\theta)\Big(\frac{\sqrt{Q}}{B\theta}\Big)^{\frac{-2+\theta}{-1+\theta}}
    \theta\left(1-\left(1+6^{\frac{\theta}{2-2\theta}}B\left(\frac{\sqrt{Q}}{B\theta}\right)^{\frac{\theta}{-1+\theta}}\right)\theta\right)\left(-c_{6}\sqrt{Q}(-3+\theta)
    \right.\right.
\end{eqnarray*}
\begin{eqnarray*}
    \left.\left.-2Q(-3+\theta)+2^{3+\frac{1}{1-\theta}}\times 3^{\frac{-2+\theta}{-1+\theta}}c^{2}Q^{\frac{1}{-1+\theta}}\alpha\kappa^{2}+2c^{2}Q(2+\gamma)(-3+\theta)\kappa^{2}-6^{\frac{-(-3+\theta)\theta}{2}}c^{2}Q^{\frac{(-2+\theta)(-1+\theta)}{2}}
    \beta(-3+\theta)\right.\right.
\end{eqnarray*}
\begin{eqnarray*}
    \left.\left.(-2+\theta)(-1+\theta)^{3}(B\theta)^{\frac{1}{-1+\theta}}\kappa^{2}-3\sqrt{6}c^{2}\sqrt{\frac{1}{Q}}\alpha(-3+\theta)(-2+\theta)(-1+\theta)(B\theta)^{\frac{2}{-1+\theta}}\kappa^{2}\right)
    +Q(1-\theta)r\right.
\end{eqnarray*}
\begin{eqnarray*}
    \left.\left(-c_{6}\sqrt{Q}(-3+\theta)+2Q(-3+\theta)
    -2^{3+\frac{1}{1-\theta}}\times3^{\frac{-2+\theta}{-1+\theta}}c^{2}Q^{\frac{1}{-1+\theta}}\alpha\kappa^{2}-2c^{2}Q(2+\gamma)(-3+
    -\theta)\kappa^{2}+2^{3+\frac{1}{1-\theta}}\times \right.\right.
\end{eqnarray*}
\begin{eqnarray*}
   \left.\left.
   3^{\frac{-2+\theta}{-1+\theta}}c^{2}Q^{\frac{1}{-1+\theta}}\alpha(-1+\theta)\kappa^{2}-2^{\frac{2+3\theta
    -\theta^{2}}{2}}\times3^{\frac{-(-3+\theta)\theta}{2}}c^{2}Q^{\frac{(-2+\theta)(-1+\theta)}{2}}\beta(-3+\theta)(-1+\theta)^{2}(B\theta)^{\frac{1}{-1+\theta}}\kappa^{2}
   \right.\right.
\end{eqnarray*}
\begin{eqnarray*}
   \left.\left. +6^{\frac{-(-3+\theta)\theta}{2}}c^{2}Q^{\frac{(-2+\theta)(-1+\theta)}{2}}
   \beta(-3+\theta)(-2+\theta)(-1+\theta)^{3}(B\theta)^{\frac{1}{-1+\theta}}\kappa^{2}+9\sqrt{6}c^{2}\sqrt{\frac{1}{Q}}\alpha(-3+\theta)(-2+\theta)(-1+\theta)\right.\right.
\end{eqnarray*}
\begin{eqnarray*}
  \left.\left.(B\theta)^{\frac{2}{-1+\theta}}\kappa^{2}+2Q(3-\theta)(1-c^{2}(2+\gamma)\kappa^{2})\right)\right]/\left[(1-\theta)
   \left\{c_{6}\sqrt{Q}(-3+\theta)+2Q(-3+\theta)-2^{3+\frac{1}{1-\theta}}
  \right.\right.
\end{eqnarray*}
\begin{eqnarray*}
    \left.\left. \times3^{\frac{-2+\theta}{-1+\theta}}c^{2}Q^{\frac{1}{-1+\theta}}\alpha\kappa^{2}-2c^{2}Q(2+\gamma)(-3+\theta)\kappa^{2}+6^{\frac{-(-3+\theta)\theta}{2}}c^{2}Q^{\frac{(-2+\theta)(-1+\theta)}{2}}
   \beta(-3+\theta)(-2+\theta)(-1+\theta)^{3}(B\theta)^{\frac{1}{-1+\theta}}\kappa^{2}\right.\right.
\end{eqnarray*}
\begin{equation}
   \left.\left.+3\sqrt{6}c^{2}\sqrt{\frac{1}{Q}}\alpha(-3+\theta)(-2+\theta)(-1+\theta)(B\theta)^{\frac{2}{-1+\theta}}\kappa^{2}\right\}\right]\geq 0.
\end{equation}

\end{itemize}

\subsection*{Case IV:}

Using the equations (\ref{Sol10}), (\ref{Sol11}), (\ref{Sol12}), (\ref{eff1}), and (\ref{eff2}), we get the following inequalities for WEC, NEC, DEC, SEC for the GO model:\\

\begin{itemize}

\item \textbf{WEC, NEC and DEC:}

\textbf{WEC and DEC}

\begin{equation*}
\left[c_{7}\sqrt{q}+q\left(1-c^{2}\kappa^{2}\left(\alpha-\frac{\beta}{\lambda}\right)\right)+\sqrt{\frac{3}{2}}c^{2}n\sqrt{q}\beta
\kappa^{2}\log{q}-q\left\{1+\frac{c_{7}}{2\sqrt{q}}+\frac{\sqrt{3}c^{2}n\beta\kappa^{2}}{\sqrt{2q}}-c^{2}\kappa^{2}\left(\alpha-\frac{\beta}{\lambda}\right)\right.\right.
\end{equation*}
\begin{equation}
\left.\left.+\frac{\sqrt{3}c^{2}n\beta\kappa^{2}\log{q}}{2\sqrt{2q}}\right\}\right]/\left[1+\frac{c_{7}}{2\sqrt{q}}+\frac{\sqrt{3}c^{2}n\beta\kappa^{2}}{\sqrt{2q}}-c^{2}\kappa^{2}\left(\alpha-\frac{\beta}{\lambda}\right)+\frac{\sqrt{3}c^{2}n\beta\kappa^{2}\log{q}}{2\sqrt{2q}}\right]\geq0
\end{equation}
and\\
\textbf{WEC and NEC}
\begin{eqnarray*}
\left[q\left\{2\left(\lambda\left(2\sqrt{q}\left(3\lambda-1+\sqrt{\frac{6}{q}}n\lambda\right)+c_{7}\left(6\lambda-1+\sqrt{\frac{6}{q}}n\lambda\right)\right)+c^{2}\kappa^{2}\left(n\beta\lambda\left(\frac{6n\lambda}{\sqrt{q}}-\sqrt{6}\right)\right.\right.\right.\right.
\end{eqnarray*}

\begin{eqnarray*}
\left.\left.\left.\left.+2\sqrt{q}\left(3\lambda-1+\sqrt{\frac{6}{q}}n\lambda\right)\left(\beta-\alpha\lambda\right)\right)\right)+c^{2}n\beta\kappa^{2}\lambda\left(-\sqrt{6}+6\left(\sqrt{6}+\frac{n}{\sqrt{q}}\right)\lambda\right)\log{q}\right\}\right]/
\end{eqnarray*}
\begin{equation}
\left[3\lambda\left(2\left(\left(c_{7}+2\sqrt{q}\right)\lambda+c^{2}\kappa^{2}\left(\sqrt{6}n\beta\lambda+2\sqrt{q}\left(\beta-\alpha\lambda\right)\right)\right)+\sqrt{6}c^{2}n\beta\kappa^{2}\lambda\log{q}\right)\right]\geq0
\end{equation}

\item \textbf{DEC:}

\begin{equation}
\left|\sqrt{2q}{3}n+q-\frac{q}{3\lambda}\right|-\frac{q\lambda\left(2c_{7}-2\sqrt{6}c^{2}n\beta\kappa^{2}+\sqrt{6}c^{2}n\beta\kappa^{2}\log{q}\right)}{2\left[\lambda\left(c_{7}+2\sqrt{q}\right)c^{2}\kappa^{2}\left(\sqrt{6}n\beta\lambda+2\sqrt{q}\left(\beta-\alpha\lambda\right)\right)\right]+\sqrt{6}c^{2}n\beta\kappa^{2}\lambda\log{q}}\leq0
\end{equation}

\item \textbf{SEC:}

\begin{eqnarray*}
\left[q\left\{2\left(\lambda\left(2\sqrt{q}\left(3\lambda-1+\sqrt{\frac{6}{q}}n\lambda\right)+c_{7}\left(4\lambda-1+\sqrt{\frac{6}{q}}n\lambda\right)\right)+c^{2}\kappa^{2}\left(n\beta\lambda\left(\frac{6n\lambda}{\sqrt{q}}-\sqrt{6}+2\sqrt{6}\lambda\right)\right.\right.\right.\right.
\end{eqnarray*}

\begin{eqnarray*}
\left.\left.\left.\left.+2\sqrt{q}\left(3\lambda-1+\sqrt{\frac{6}{q}}n\lambda\right)\left(\beta-\alpha\lambda\right)\right)\right)+c^{2}n\beta\kappa^{2}\lambda\left(-\sqrt{6}+4\sqrt{6}\lambda+6\frac{n}{\sqrt{q}}\lambda\right)\log{q}\right\}\right]/
\end{eqnarray*}
\begin{equation}
\left[\lambda\left(2\left(\left(c_{7}+2\sqrt{q}\right)\lambda+c^{2}\kappa^{2}\left(\sqrt{6}n\beta\lambda+2\sqrt{q}\left(\beta-\alpha\lambda\right)\right)\right)+\sqrt{6}c^{2}n\beta\kappa^{2}\lambda\log{q}\right)\right]\geq0
\end{equation}

The corresponding expressions for the Chen-Jing model read as,\\

\item\textbf{WEC, NEC and DEC:}

\textbf{WEC and DEC}

\begin{eqnarray*}
\left[c_{8}\sqrt{q}-18n^{2}\alpha\kappa^{2}+q\left(1-c^{2}\kappa^{2}\left(\gamma-\frac{2\alpha}{\lambda^{2}}+\frac{4\alpha}{\lambda}-\frac{\beta}{\lambda}\right)\right)-\sqrt{3/2}c^{2}n\sqrt{q}\kappa^{2}\left(\beta-4\alpha+\frac{5\alpha}{\lambda}\right)\log{q}\right.
\end{eqnarray*}
\begin{eqnarray*}
\left.-q\left\{1+\frac{c_{8}}{2\sqrt{q}}-\sqrt{\frac{3}{2q}}c^{2}n\kappa^{2}\left(\beta-4\alpha+\frac{5\alpha}{\lambda}\right)-c^{2}\kappa^{2}\left(\gamma-\frac{2\alpha}{\lambda^2}+\frac{4\alpha}{\lambda}-\frac{\beta}{\lambda}\right)-\frac{1}{2}\sqrt{\frac{3}{2q}}c^{2}n\kappa^{2}\left(\beta-4\alpha+\frac{5\alpha}{\lambda}\right)\log{q}\right\}\right]/
\end{eqnarray*}
\begin{equation}
\left[1+\frac{c_{8}}{2\sqrt{q}}-\sqrt{\frac{3}{2q}}c^{2}n\kappa^{2}\left(\beta-4\alpha+\frac{5\alpha}{\lambda}\right)-c^{2}\kappa^{2}\left(\gamma-\frac{2\alpha}{\lambda^{2}}+\frac{4\alpha}{\lambda}-\frac{\beta}{\lambda}\right)-\sqrt{\frac{1}{2}\frac{3}{2q}}c^{2}n\kappa^{2}\left(\beta-4\alpha+\frac{5\alpha}{\lambda}\right)\log{q}\right]\geq0
\end{equation}
and\\

\textbf{WEC and NEC}

\begin{eqnarray*}
\left[\sqrt{q}\left\{2\left(\lambda^{2}\left(c_{8}\sqrt{q}\left(1-\left(6-\sqrt{6/q}n\right)\lambda\right)+2\left(q-3q\lambda+n\left(-\sqrt{6q}+54n\alpha\kappa^{2}\right)\lambda\right)\right)+c^{2}\kappa^{2}\left(n\lambda\sqrt{q}\left(-\sqrt{6}+\frac{6n\lambda}{\sqrt{q}}\right)\right.\right.\right.\right.
\end{eqnarray*}
\begin{eqnarray*}
\left.\left.\left.\left.\left(\alpha\left(5-4\lambda\right)+\beta\lambda\right)+2\sqrt{6q}n\lambda\left(\alpha\left(4\lambda-2\right)+\lambda\left(\gamma\lambda-\beta\right)\right)+2q\left(3\lambda-1\right)\left(\alpha\left(4\lambda-2\right)+\lambda\left(\gamma\lambda-\beta\right)\right)\right)\right)+c^{2}n\sqrt{q}\kappa^{2}\lambda\times\right.\right.
\end{eqnarray*}
\begin{eqnarray*}
\left.\left.\left(\sqrt{6}+6\left(\sqrt{6}+n\sqrt{1/q}\right)\lambda\right)\left(\alpha\left(5-4\lambda\right)+\beta\lambda\right)\log{q}\right\}\right]/\left[3\lambda\left\{2\left(\left(c_{8}+2\sqrt{q}\right)\lambda^{2}+c^{2}\kappa^{2}\left(\sqrt{6}n\lambda\left(5\alpha-4\alpha\lambda+\beta\lambda\right)\right.\right.\right.\right.
\end{eqnarray*}

\begin{equation}
\left.\left.\left.\left.+2\sqrt{q}\left(-2\alpha+4\alpha\lambda\beta\lambda+\gamma\lambda^{2}\right)\right)\right)+\sqrt{6}c^{2}n\kappa^{2}\lambda\left(\alpha\left(5-4\lambda\right)+\beta\lambda\right)\log{q}\right\}\right]\geq0
\end{equation}

\item \textbf{DEC:}

\begin{eqnarray*}
\left[-2c_{8}q\lambda^{2}+72n^{2}\sqrt{q}\alpha\kappa^{2}\lambda^{2}-2\sqrt{6}c^{2}nq\kappa^{2}\lambda\left(\alpha\left(5-4\lambda\right)+\beta\lambda\right)+\sqrt{6}c^{2}nq\kappa^{2}\lambda\left(\alpha\left(5-4\lambda\right)+\beta\lambda\right)\log{q}\right]/
\end{eqnarray*}
\begin{eqnarray*}
\left[2\left[\left(c_{8}+2\sqrt{q}\right)\lambda^{2}-c^{2}\kappa^{2}\left(\sqrt{6}n\lambda\left(5\alpha-4\alpha\lambda+\beta\lambda\right)+2\sqrt{q}\left(-2\alpha+4\alpha\lambda-\beta\lambda+\gamma\lambda^{2}\right)\right)\right]\right.
\end{eqnarray*}
\begin{equation}
\left.-\sqrt{6}c^{2}n\kappa^{2}\lambda\left(\alpha\left(5-4\lambda\right)+\beta\lambda\right)\log{q}\right]+\left|\sqrt{\frac{2q}{3}}n+q-\frac{q}{3\lambda}\right|\leq0
\end{equation}

\item\textbf{SEC:}

\begin{eqnarray*}
\left[\sqrt{q}\left\{2\left(\lambda^{2}\left(c_{8}\sqrt{q}\left(1-\left(4+\sqrt{6/q}n\right)\lambda\right)+2\left(q-3q\lambda+n\left(-\sqrt{6q}+18n\alpha\kappa^{2}\right)\lambda\right)\right)+c^{2}\kappa^{2}\left(n\lambda\sqrt{q}\left(-\sqrt{6}+\right.\right.\right.\right.\right.
\end{eqnarray*}
\begin{eqnarray*}
\left.\left.\left.\left.\left.2\sqrt{6}\lambda+\frac{6n\lambda}{\sqrt{q}}\right)\left(\alpha\left(5-4\lambda\right)+\beta\lambda\right)+2\sqrt{6q}n\lambda\left(\alpha\left(4\lambda-2\right)+\lambda\left(\gamma\lambda-\beta\right)\right)+2q\left(3\lambda-1\right)\left(\alpha\left(4\lambda-2\right)+\lambda\left(\gamma\lambda-\beta\right)\right)\right)\right)\right.\right.
\end{eqnarray*}
\begin{eqnarray*}
\left.\left.+c^{2}n\sqrt{q}\kappa^{2}\lambda\left(-\sqrt{6}+4\sqrt{6}\lambda+6n\sqrt{1/q}\lambda\right)\left(\alpha\left(5-4\lambda\right)+\beta\lambda\right)\log{q}\right\}\right]/\left[\lambda\left\{2\left(-\left(c_{8}+2\sqrt{q}\right)\lambda^{2}+c^{2}\kappa^{2}\left(\sqrt{6}n\lambda\left(5\alpha\right.\right.\right.\right.\right.
\end{eqnarray*}
\begin{equation}
\left.\left.\left.\left.\left.-4\alpha\lambda+\beta\lambda\right)+2\sqrt{q}\left(-2\alpha+4\alpha\lambda-\beta\lambda+\gamma\lambda^{2}\right)\right)\right)+\sqrt{6}c^{2}n\kappa^{2}\lambda\left(\alpha\left(5-4\lambda\right)+\beta\lambda\right)\log{q}\right\}\right]\geq0
\end{equation}

\end{itemize}

%%%%%%%%%%%%%%%%%%%%%%%%%%
\section*{Acknowledgments}
%%%%%%%%%%%%%%%%%%%%%%%%%%

PS acknowledges Government of West Bengal Science and Technology and Biotechnology
Department, for financial support to carry out Research project No.: STBT-
11012(31)/5/2024-WBSCST SEC. P.R. acknowledges the Inter-University Centre for Astronomy and Astrophysics (IUCAA), Pune, India for granting visiting
associateship. We thank the anonymous referee/referees for his/her/their invaluable comments that helped us to improve the quality of the manuscript.

\section*{Data Availability Statement}

No data was generated or analyzed in this study.

\section*{Conflict of Interest}

There are no conflicts of interest.

\section*{Funding Statement}

There is no funding to report for this article.

%%%%%%%%%%%%%%%%%%%%%%%%%%%%%%%%%%%%%%%%%%%%%%%%%%%%%%%%%%%%%%%%%%%%%
%%%%%%%%%%%%%%%%%%%%%%%%%%%%%%%%%%%%%%%%%%%%%%%%%%%%%%%%%%%%%%%%%5

\end{document}